\documentclass[
  journal=pasa,
  manuscript=research-paper, 
  year=2025,
  volume=YY,
]{cup-journal}

\usepackage[utf8]{inputenc}
\usepackage{amssymb}
\usepackage{multirow}
\usepackage{xcolor}
\usepackage{cancel}
\usepackage[normalem]{ulem}
\usepackage{amsmath}
\usepackage{mathtools}
\usepackage[nopatch]{microtype}
\usepackage{siunitx}
\usepackage{booktabs}
\usepackage{physics}
\usepackage{gensymb}
\usepackage{hyperref}
\usepackage{microtype}
\usepackage{tabularray}

\usepackage{aas-macros}
\received {dd Mmm YYYY}
\revised  {dd Mmm YYYY}
\accepted {dd Mmm YYYY}
\published{22 September 202X}

\title{Determining Ideal Fields for Epoch of Reionisation Science Using the 21 cm Line}

\author{E. Jong}
\affiliation{International Centre for Radio Astronomy Research, Curtin University, Bentley WA, Australia}
\affiliation{ARC Centre of Excellence for All-Sky Astrophysics in 3D (ASTRO 3D), Australia}
\email[E. Jong]{eric.jong@postgrad.curtin.edu.au}

\author{C.~M. Trott}
\affiliation{International Centre for Radio Astronomy Research, Curtin University, Bentley WA, Australia}
\affiliation{ARC Centre of Excellence for All-Sky Astrophysics in 3D (ASTRO 3D), Australia}

\author{C.~D. Nunhokee}
\affiliation{International Centre for Radio Astronomy Research, Curtin University, Bentley WA, Australia}
\affiliation{ARC Centre of Excellence for All-Sky Astrophysics in 3D (ASTRO 3D), Australia}

\author{Q. Zheng}
\affiliation{Shanghai Astronomical Observatory, Chinese Academy of Sciences, 80 Nandan Road,
Shanghai, 200030, P. R. China}


\begin{document}

\begin{abstract}
    The upcoming Square Kilometre Array Low Frequency (SKA-Low) interferometer will have the
    required sensitivity to detect the 21 cm line from neutral hydrogen during the Epoch of
    Reionisation (EoR). In preparation, we investigated the suitability of different fields for EoR
    science with the 21~cm line, using existing observations of candidate fields from the Murchison
    Widefield Array (MWA). Various image and calibration metrics were extracted from archival MWA
    observations centred on $z \sim 6.8$.  We explore the usefulness of these metrics and compare
    their behaviour between different fields of interest. In addition, a theoretical approach to
    quantifying the impact of different fields on the power spectrum is also provided. Gain
    uncertainties were calculated based on the positions of the calibrators within the beam. These
    uncertainties were then propagated into visibilities to produce cylindrical power spectra for
    various fields. Using these metrics in combination with the power spectra, we confirm that EoR0
    ($\text{R.A.} = \SI{0}{\deg}$, $\text{Dec} = \SI{-27.0}{\deg}$) is an ideal EoR field and
    discuss the interesting behaviour of other fields.
\end{abstract}

\section{Introduction}
From the Cosmic Dawn (CD, $z \sim 30-10$) through the Epoch of Reionisation (EoR, $z \sim 10-5.3$), the
Universe underwent incredible change. During the CD, neutral hydrogen (HI) atoms that formed at
Recombination start to collapse gravitationally and form the first luminous objects. These objects
then re-ionised the HI that surrounded them and created ``bubbles'' of ionised hydrogen (HII) --- the
beginning of EoR.

Over time, with the formation of new luminous sources and bubbles continuously expanding, we arrive
at the bright and mostly ionised Universe of the present day. Understanding the evolution of
hydrogen helps to reveal the properties of the first stars, such as the processes involved in the
formation of these stars, their mass, when they first formed, and how the Universe evolved from a
smooth matter distribution to its complex structures of today (\cite{Furlanetto2006,
Zheng2020, Koopmans2015}). Evidently, there lies a wealth of knowledge crucial to our understanding
of the Universe. 

To trace neutral hydrogen, the 21 cm wavelength photon produced by the hyperfine transition in the
ground state of hydrogen is of interest. This particular wavelength is not easily re-absorbed by
other hydrogen atoms, hence it travels through the neutral hydrogen medium unimpeded. Although it
may take multiple millions of years for a single hydrogen atom to produce this signal
(\cite{Storey1994}), the abundance of hydrogen ensures that we can probe these time periods. For a
thorough review of the history of HI in the Universe, see \cite{Pritchard2012, Koopmans2015,
Furlanetto2006} and references therein.

Major efforts in EoR science lie in measuring the brightness temperature of the
$\SI{21}{\centi\metre}$ line at some particular redshift, through a variety of statistics; the
brightness temperature spatial power spectrum is commonly used. This is undertaken through
observations with radio interferometers such as the Murchison Widefield Array (MWA,
\cite{Tingay2013, Wayth2018}), LOw-Frequency Array (LOFAR, \cite{Haarlem2013}), New Extension in
Nan\c{c}ay Upgrading LOFAR (NENUFAR, \cite{Zarka2015}), Hydrogen Epoch of Reionization Array (HERA,
\cite{DeBoer2017}), the Giant Metre-wave Radio Telescope (GMRT,
\cite{e324b39f-8469-3fce-bd35-bf9f32836148}), and the Long Wavelength Array (LWA,
\cite{Ellingson2009}). Results for the upper-limit on the $\SI{21}{\centi\metre}$ brightness
temperature have also been reported by these instruments: MWA (\cite{Trott2020, Nunhokee2025}),
LOFAR (\cite{Mertens2020, Acharya2024, Mertens2025}), NenuFAR (\cite{Munshi2024, Munshi2025}), HERA
(\cite{Abdurashidova2022, HERACollaboration2023}), GMRT (\cite{Paciga2011}), and LWA
(\cite{Eastwood2019}).

There are many challenges in the measurement of the $\SI{21}{\centi\metre}$ signal. Radio signals from astrophysical
and human sources produce foreground contamination; processes such as synchrotron emission, free-free
scattering, bright radio sources, radio emission from digital television channels, FM radio emissions and
satellites are orders of magnitude brighter than the $\SI{21}{\centi\metre}$ line (\cite{Bowman2010, Chapman2019,
Offringa2015}). Imprecise calibration solutions have also been a challenge in these experiments.
Spectral features in the calibration solutions can propagate into the power spectra and affect our
ability to make a measurement.
Spectral features caused by using an incomplete sky model in the calibration step (\cite{Barry2016,
EwallWice2017, Procopio2017}) or errors in beam responses (\cite{Nunhokee2020, Chokshi2024,
Brackenhoff2025}) have been shown to overwhelm the feeble $\SI{21}{\centi\meter}$ signal. Clearly,
the study of these processes is crucial to minimising their effect and for the success of EoR
science.

This paper aims to investigate one such challenge, that is, how different parts of the sky impact
data calibration and in turn determine the ideal fields of the sky for EoR observations with
SKA-Low. The calibration is sensitive to different parts of the sky due to the density of sources,
brightness distribution of sources, and the types of sources present (compact or extended).

The MWA collaboration has selected two regions of the sky which are deemed fit for EoR science: EoR0
($\text{R.A.} = \SI{0}{\degree}$, $\text{Dec}=\SI{-27}{\degree}$), and EoR1($\text{R.A.} =
\SI{60}{\degree}$, $\text{Dec}=\SI{-27}{\degree})$. EoR0 is described as containing a few, bright
resolved sources allowing for ``easier'' calibration (\cite{Jacobs2016}). EoR1 was
chosen for similar reasons, containing another cold patch of the sky.

The LOFAR collaboration observes two main fields, the North Celestial Pole (NCP,
\cite{Yatawatta2013}), and 3C196 (\cite{Asad2015, Ceccotti2025}). The NCP was selected (apart from the
benefits due to the location of the interferometer) for its position in a relatively cold spot of
the galactic halo, which reduces foregrounds, and because it did not contain an extremely bright
source, which results in less artefacts from deconvolution. The 3C196 field was chosen due to its
position in a colder region of the galactic halo and due to the presence of a very bright unresolved
source located near the centre resulting in accurate direction-independent calibration. 

The HERA collaboration, which uses a zenith pointed array, observes various fields throughout a
large range of LSTs. Some fields were selected due to minimal diffuse foregrounds and the presence
of a bright source from the GLEAM catalogue (\cite{Wayth2015, Abdurashidova2022}). Additional fields
were selected to avoid Fornax A and the galactic centre (\cite{HERACollaboration2023}).

More recently, candidate fields have been chosen by \cite{Zheng2020} for the upcoming SKA-Low, using a
limited number of parameters to determine field quality, such as minimal galactic emission chosen by
cool regions of the Haslam $\SI{408}{\mega\hertz}$ All-Sky map (\cite{1982A&AS...47....1H}), far from the
Magellanic clouds, contain minimal bright radio sources that are uniformly positioned within the
beam, and should not contain resolved radio diffuse sources. These parameters help to select fields
with the intention of making the subsequent post-processing steps (foreground subtraction,
calibration, and imaging) easier, and more accurate. Further processing and analysis of one
of the candidate fields from this work, and several other ``quiet'' fields show that selecting a
suitable field is critical for EoR science with the SKA (Zheng et al. in prep.).

In this work we extend those parameters, and also extend the number of fields studied. We take both
a data-driven and a theoretical approach to this problem to determine the key parameters for
determining the quality of a field for EoR science. We use observations of existing fields from the
MWA telescope and study the actual calibration precision. We also apply a theoretical approach to
predict the most precise calibration that can be obtained for each field.
This paper will focus on MWA parameters for characterising fields, but the applicability to the
SKA-Low will also be discussed.

This paper is structured as follows, Section \ref{data_methods} goes through the details and process
in obtaining data used in this work. Section \ref{metrics} will discuss the theoretical aspects and
desired behaviour of the chosen metrics in this work. Section \ref{theoretical_method} discusses the
theory behind calculating the gain uncertainties. Section \ref{computational_method} describes the
computational aspects of the work. In Section \ref{results} we present the results for the extracted
metrics and the theoretical gain uncertainties. In Section \ref{discussion} we discuss and compare
the results for the different fields, and finally we conclude the work in Section \ref{conclusion}.

\section{Data method}
\label{data_methods}

\subsection{Murchison Widefield Array details}
The MWA is a radio interferometer located at the Murchison Radio-Astronomy Observatory, a radio quiet
zone in Western Australia (\cite{Tingay2013}). In this work we are concerned with the phase II
extended configuration (\cite{Wayth2018}) of the telescope. This configuration contains 128 tiles
with a maximum baseline of approximately $\SI{5.2}{\kilo\meter}$. Its operating frequencies are from
$\SI{70}{\mega\hertz}$ to $\SI{300}{\mega\hertz}$, with a field of view of approximately 26 degrees
at $\SI{150}{\mega\hertz}$. The MWA produces data at a spectral resolution of
$\SI{10}{\kilo\hertz}$, and a temporal resolution of 0.5 seconds. Each observation is typically 2
minutes in duration covering only a $\SI{30.72}{\mega\hertz}$ instantaneous bandwidth.

\subsection{Data collection and pre-processing}
Data were obtained from the MWA ASVO archive \footnote{https://asvo.mwatelescope.org/} through the
MWA TAP service. The TAP service was queried with the following settings: observations are within a
circle of radius $5\degree$ of the phase centre, and centred on channel 144 (184.32 MHz, which corresponds
to $z \sim 6.8$). With these settings, we obtain observations centred on one of two frequency bands
designated for EoR. Furthermore, the settings allow us to obtain as many observations as we can
without the sky changing significantly between observations. As mentioned earlier, these
observations were made in the Phase II configuration, which allows us to extend upon EoR
observations performed by \cite{Zheng2020}.

\begin{figure*}
    \begin{center}
        \includegraphics[width=0.95\textwidth]{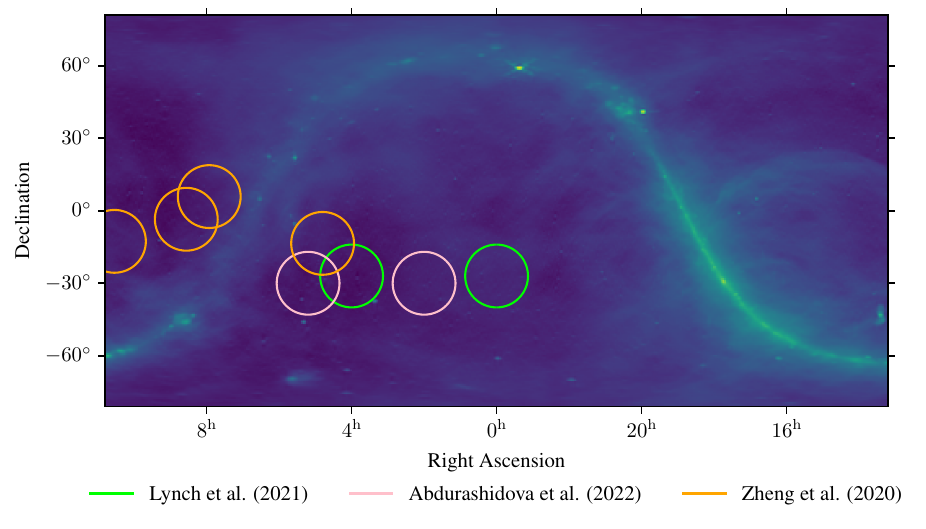}
    \end{center}
    \caption{Fields of interest in this study highlighted on the $\SI{408}{\mega\hertz}$ all-sky map
        (\cite{1982A&AS...47....1H}). Highlighted in green are the MWA EoR1 and EoR2 fields
        (\cite{Lynch2021}). In pink are HERA fields (\cite{Abdurashidova2022}). In orange are fields
        selected by other metrics (\cite{Zheng2020}).}\label{fig:fields_on_sky}
\end{figure*}

In this work, we inspect the fields used by MWA (EoR0 and EoR1, \cite{Barry2019,
Trott2020, Rahimi2021}), fields which have been used by HERA (\cite{Abdurashidova2022}),
and fields which were chosen by other metrics (prefixed with ``SKAEOR'', \cite{Zheng2020}). The phase centres and
number of observations for all the fields investigated are given in Table \ref{tab:fields_tab}. A
figure highlighting the fields on the radio sky is given in Figure \ref{fig:fields_on_sky}. Due to the use
of analog beamformers in the MWA, the interferometer is only capable of coarse pointing
(\cite{Tingay2013}). Thus, from the perspective of the MWA the sky is broken into certain ``grid
numbers'' (also known as pointing numbers) which describe the pointing. For each field, numerous
observations are used, in order to separate field-based structure from poor observations (e.g., due
to the health of the telescope itself).

\begin{table}[h]
    \caption{Table of fields which have been downloaded and processed, alongside the right ascension
    (deg) and declination (deg) of their phase centres. Listed are EoR0 and EoR1 (\cite{Lynch2021}),
fields used by HERA (\cite{Abdurashidova2022}), and fields chosen by other metrics (prefixed with
``SKAEOR'', \cite{Zheng2020})}\label{tab:fields_tab}
    \begin{center}
        \begin{tabular}[c]{c|c|c|c}
            \hline
            \multicolumn{1}{c|}{\textbf{Field name}} & 
            \multicolumn{1}{c|}{\textbf{RA (deg)}} &
            \multicolumn{1}{c|}{\textbf{DEC (deg)}} &
            \multicolumn{1}{c}{\textbf{Num. obs.}} \\
            \hline
            EoR0 & 0.00 & -30.00 & 55 \\ 
            EoR1 & 60.00 & -30.00 & 35 \\ 
            HERA LST 2.0 & 30.00 & -30.00 & 31 \\
            HERA LST 5.2 & 78.00 & -30.00 & 86 \\
            SKAEOR5 & 118.91 & 5.86 & 100 \\ 
            SKAEOR6 & 128.40 & -3.52 & 26 \\ 
            SKAEOR14 & 158.14 & -12.66 & 53 \\ 
            SKAEOR15 & 72.5 & -13.35 & 100 \\ 
            \hline
        \end{tabular}
    \end{center}
\end{table}

Once the data had been downloaded, they were calibrated for direction-independent gains with
Hyperdrive\footnote{\url{https://github.com/MWATelescope/mwa_hyperdrive}} (\cite{Jordan2025}). The
sky model used is a combination of different surveys and models, including LoBES (\cite{Lynch2021})
for the EoR 0 field, \cite{Procopio2017} for the EoR1 field, \cite{Line2020} for a shapelets model
of Fornax A, (\cite{Cook2022}) for the Centaurus A model and Galactic Plane Supernova Remnants, and
the GLEAM survey (\cite{Wayth2015, HurleyWalker2016}). Only Stokes I information were used for the
sky model. We used the brightest 8000 sources. In addition, Hyperdrive uses a per-channel
calibration at the resolution of $\SI{2}{\second}$/$\SI{40}{\kilo\hertz}$. Furthermore, Hyperdrive
considers the leakage terms in the Jones matrix of the beam response, utilises a simulated Full
Embedded Element (FEE) beam model (\cite{Sokolowski2017}), and solves for calibration solutions per
frequency channel using a similar process described by \cite{Mitchell2008}. No antenna flagging
algorithms were applied to the data.

Calibration solutions were assessed using the MWA's Quality Analysis pipeline (\cite{Nunhokee2024}).
Amplitude calibration solutions were normalised to the median and the phase solutions were also
unwrapped before calculating the metrics.

The calibrated visibilities were imaged with WSCLEAN (\cite{Offringa2014}). Deconvolution (CLEAN,
\cite{Hoegbom1974}) was applied until the data reached the $3\sigma$ noise threshold or reached the
maximum number of iterations (set to 10000). Since the field of view of the main lobe of the MWA
beam is approximately 21 degrees at 184.32 MHz, we set scale of each pixel to 15 arcseconds and the
image size to $5064 \times 5064$ pixels.

\section{Metrics}
\label{metrics}

This work aims to provide a data driven approach to selecting ideal fields for EoR science. With
this motivation, we will explore the insight and usefulness of a variety of metrics in both the image
and calibration space. The metrics to be discussed and their desired behaviour are given in
Table \ref{tab:metrics}.

\begin{table}
    \caption{Table of metrics to be extracted from archival MWA data, and their desired
    behaviour.}\label{tab:metrics}
    \begin{center}
        \begin{tabular}[c]{l|l}
            \hline
            \multicolumn{1}{c|}{\textbf{Metric}} & 
            \multicolumn{1}{c}{\textbf{Desired behaviour}} \\
            \hline
            Image root-mean-square noise & Low \\
            Image dynamic range & High \\
            Calibration Gain smoothness & Low \\
            Calibration Phase RMSE & Low \\
            Calibration Phase MAD & Low \\
            Calibration Phase Euclidean distance & Low \\
            Calibration Phase KS-metric & Low \\
            \hline
        \end{tabular}
    \end{center}
\end{table}

The following sections will first have a brief explanation for the motivation and expected
behaviour, followed by more detailed discussion.

\subsection{Image Metrics}
\textbf{Motivation:} Residual signal in peeled images indicates the presence of
unmodelled source sidelobe.\\ \textbf{Expected behaviour:} Root-mean-square noise should be
minimised and dynamic range maximised.\\

In this work, deconvolved multi-frequency-synthesis images are produced. Hence, residual signal in
these images indicates the presence of unmodelled source sidelobes. The root-mean-square (RMS) and
dynamic range (DR) are used to quantify the residual signal. The RMS should be minimised and DR
maximised; a high RMS may completely obscure the $\SI{21}{\centi\metre}$ signal and may indicate
significant contamination from source sidelobes, while a low DR indicates a lack of bright
calibrators or, again, contamination from source sidelobes. The RMS and DR are calculated with the
following equations

\begin{equation}
    \sigma_\text{rms} = \sqrt{\frac{1}{N}\sum_i^N{x_i^2}}
    \label{eq:rms}
\end{equation}

\noindent and

\begin{equation}
    \chi = \frac{X}{\sigma_\text{rms}}
    \label{eq:dr}
\end{equation}

\noindent respectively, where $x_i$ is the value of each pixel, $N$ is the number of pixels, and $X$
is the value of the brightest pixel.

Deconvolving in the image plane is conceptually related to the process of peeling in the visibility
plane: they both aim to remove unwanted sidelobes from the data. The peeling of visibilities is a
common step in creating power spectra for EoR science, hence the RMS of deconvolved images can still
be suitable for determining candidate EoR fields. However, in this work, the RMS metric is intended
as a simple diagnostic to compare different fields. Therefore, it is most important to ensure
consistency in how these images are generated.

\subsection{Amplitude solutions smoothness}
\textbf{Motivation:} Spectral features in gain amplitudes propagate into final
statistics of interest. \\
\textbf{Expected behaviour:} Smoothness metric should be minimised.\\

In the calibration solutions case, we investigate the amplitude and phase of the calibration
solutions separately. For the amplitude solutions we inspect the smoothness of the solutions across
frequency. The smoothness is especially important since Hyperdrive, our calibration tool, makes no
assumptions on how the complex gains should behave; Hyperdrive solves for solutions per-frequency,
hence capturing most of the spectral features. In methods that use regularisation, assumptions about
how the gains behave are made, and a penalty function is used to constrain the solutions
(\cite{Ikeda2025, Yatawatta2015}) which may bias the solutions towards a certain class of solutions.

Amplitude solutions that are not smooth --- which may be the result of the particular observing
field or an error at the time of observation --- can contain spectral structure due to a poor sky
model which will propagate into the final power spectrum (\cite{Byrne2019, EwallWice2017,
Barry2016}). This can heavily affect the ability to make a measurement, hence we suggest that smooth
amplitude solutions are ideal. We calculate the smoothness of the amplitude solutions with the
following


\begin{equation}
    S = \frac{1}{M} \displaystyle\sum_{i = 1}^{N_+} m_i,
    \label{eq:smoothness}
\end{equation}

\noindent where $M$ is the value of the DC-mode of the Fourier Transform (FT) of the amplitude
solutions, $m_i$ is the value of the $i$-th positive Fourier mode (after normalising to the median),
and $N_+$ is the number of positive Fourier modes. Since the amplitude solutions are not a periodic
signal, we also apply the Blackman-Harris window to reduce spectral leakage (\cite{Lessard2006}).

\subsection{Phase solutions metrics}
\textbf{Motivation:} Phase solutions should be linear and similar between the two
polarisations.\\
\textbf{Expected behaviour:} Root-mean-square-error, mean-absolute-deviation, KS-test, and Average
Euclidean distance should all be minimised.\\

Hyperdrive applies direction-independent calibration, meaning it calibrates for time delays in the
system which arise from factors such as cable lengths from antenna to receiver. A time delay
in the signal corresponds to a linear phase shift in Fourier space. Hence, we investigate the
linearity of the phase solutions with the root-mean-square-error (RMSE) and mean-absolute-deviation
(MAD). The RMSE and MAD are calculated using the following

\begin{equation}
    \text{RMSE} = \sqrt{\frac{1}{N}\sum_1^N{(x_i - \hat{x}_i)^2}}
    \label{eq:rmse}
\end{equation}

\noindent and

\begin{equation}
    \text{MAD} = \frac{1}{N} \sum_1^{N} \abs{r_i - R},
    \label{eq:mad}
\end{equation}

\noindent where $N$ is the number of frequency channels, $x_i$ is the antenna's phase solution at
frequency channel $i$, $\hat{x_i}$ is the phase solution predicted by the line of best fit, $r_i$
are the residuals between the observed and predicted phase solutions and $R$ is the mean value of
the residuals.

Another intriguing aspect of the phase solutions arise from the fact that the NS and EW dipoles have
differing response patterns.  Hence, investigating the similarity between the NS and EW polarisation
phase solutions may provide interesting insight in the behaviour of a field. Here, we suggest
solutions which are similar to each other are better than those which are not similar. A
dissimilarity can indicate bright sources are present in one polarisation and not the other, or
suggest an error occurred during observation in one polarisation. We use two metrics to describe the
similarity, the average Euclidean distance between the two sets of solutions, and the
Kolgomorov-Smirnov (KS) metric. The Euclidean distance is simply the average of the distance between
the NS and EW phase solutions at each channel. A smaller value indicates that the solutions are
closer. The KS metric tests how similar the underlying empirical distribution functions of the two
sets of solutions are, a value of 0 indicates the two sets are identical while a value of 1
indicates they are the most dissimilar.


\section{Theoretical method}
\label{theoretical_method}
In addition to the metrics, a theoretical approach was used to assess the information content of
data from different fields and the consequent impact on calibration precision.


\subsection{Gain uncertainties}
For the theoretical analysis of gain uncertainties, we have used the Crame\'r-Rao bound (CRB)
statistic. The CRB provides a lower bound on the variance of an unbiased estimator; or,
equivalently, it provides the precision of an unbiased estimator. It does not provide a method for
estimating the unknown parameters. The CRB has been used previously to investigate the effects of
imprecise calibrator parameters such as the position, spectral index, and brightness
(\cite{Trott2011}). In this work, the same methods were applied to investigate the best a
calibration pipeline can estimate gain uncertainties.

We begin with the Fisher Information Matrix, an element of which can be written in the form
(\cite{Kay1993}) 

\begin{equation}
    I_{ab}(\boldsymbol{\theta}) = \frac{2}{\sigma^2} \int{ \frac{\partial s^{H}(t;
    \boldsymbol{\theta})}{\partial\theta_a}\frac{\partial
s(t;\boldsymbol{\theta})}{\partial\theta_b} dt},
    \label{eq:simplified_FIM}
\end{equation}

\noindent where $\sigma^2$ is the variance, $s(t; \boldsymbol{\theta})$ is a signal,
$H$ denotes the Hermitian conjugate, and $\boldsymbol{\theta}$ is a vector of
unknown parameters which the signal is conditioned on. The signal is given by

\begin{equation}
    V_{ab} = \sum_{j=1}^{N_c} B_j g_a \overline{g_b} \text{exp}(-2 \pi i(u_{ab}l_j + v_{ab}m_j)),
    \label{eq:vis}
\end{equation}

\noindent where $g_a$ and $g_b$ are the complex gains of antennas $a$ and $b$
respectively, $u_{ab}$ and $v_{ab}$ are the baseline coordinates between antennas $a$ and $b$, $B_j$
is the brightness of calibrator $j$, and $l_j$ and $m_j$ are the calibrator's coordinates (direction
cosines). In this study, we have assumed $g_a = g_b = 1$, i.e. unity gains.

We are concerned with how well we can estimate the gains $g_a$ and $g_b$, hence for a telescope with
$N$ antennas we have $\boldsymbol{\theta} = [g_1, g_2, g_3, \dots, g_{ N }]$. This
means $s(t; \boldsymbol{\theta})$ in Equation \ref{eq:simplified_FIM} is a vector of visibilities
measured by every baseline. Each element of this vector of length $N_{\text{baselines}}$ is given by
Equation \ref{eq:vis}, where $N_{\text{baselines}} = N^2$ is the number of baselines, including
conjugates, and auto-correlations to complete the matrix. Taking the partial derivative of this
vector with respect to $g_a$ we have

\begin{equation}
    \frac{\partial s_{ab}(t; \boldsymbol{\theta})}{\partial g_a} = \sum_{j=1}^{N_c} B_j
        \overline{g_b} \text{exp}(-2 \pi i(u_{ab}l_j + v_{ab}m_j)).
    \label{eq:ds_vec}
\end{equation}

This means most of the vector, $\frac{\delta \boldsymbol{s}(t;\boldsymbol{\theta})}{\delta g_a}$,
will possess a value of 0 and $N$ elements will be populated, since $N_\text{baselines} - N$
elements do not contain $g_a$ and will be differentiated to 0. 

When $a \neq b$, the product of the two partial derivatives is then only non-zero
when the product contains $g_a$ and $g_b$. For example, for antennas 1 and 2 the product is non-zero
when the terms $g_1$ and $g_2$ are both present in the product. When $a = b$, the
product of the partial derivatives is non-zero at $N$ elements, $N -1$ of those elements are
identical, and 1 element will have a multiplicative factor of 4 due to the partial derivative of a $g_a^2$
term. Hence, when $a = b$, an additional multiplicative factor of $N + 3$ is
needed, this condition can be expressed with a Kronecker-delta function. The final expression for an
element of the Fisher Information Matrix is

\begin{align}
    &I_{ab}(\boldsymbol{\theta}) = (N + 3)^{\delta_{a b}} \times \frac{2}{\sigma^2} \times \nonumber\\
    &\int{\displaystyle\sum_{j = 1}^{N_c}\displaystyle\sum_{k = 1}^{N_c}B_j B_k g_a \overline{g_b} \exp{-2\pi
    i(u_{ab}(l_j - l_k) + v_{ab}(m_j - m_k)} dt}.
    \label{eq:final_FIM}
\end{align}

The variance $\sigma^2$ of a single visibility signal measured by a single baseline is given by

\begin{equation}
    \sigma = \displaystyle\frac{2k_b T_\text{sys}}{A_\text{eff} \sqrt{\Delta \nu \Delta \tau}},
    \label{eq:radiometer_full}
\end{equation}

\noindent where $k_b$ is the Boltzmann constant, $T_\text{sys}$ is the system temperature,
$A_\text{eff}$ is the effective area of an antenna, $\Delta \nu$ is the bandwidth of a single
measurement, and $\Delta \tau$ is the integration time for a single measurement. 

The CRB matrix is then the inverse of Equation \ref{eq:final_FIM}, and the gain uncertainties can be
found on the square-root of the diagonal of this matrix. Previous work with the CRB
(\cite{Trott2011}) has shown one bright calibrator produces better gain and phase precision compared
to many lower brightness sources.

\subsection{Propagation of gain uncertainties}
To propagate gain uncertainties into the visibilities we use the standard covariance matrix method
given by

\begin{equation}
    \sigma_{V_{ab}}^2 = \left[\mathbf{J} \mathbf{C}_{\theta} \mathbf{J}^\dagger \right]_{ab},
    \label{eq:covar_method}
\end{equation}

\noindent where $\mathbf{J}$ is the Jacobian of partial derivatives of the visibility function with
respect to the parameters $\theta_{ab} = (g_{a}, g_{b})$, $\mathbf{J}^\dagger$ is the complex
conjugate, and $\mathbf{C}_{\theta}$ is the covariance matrix of parameter uncertainties. Expanding
out equation \ref{eq:covar_method}, we arrive at the following

\begin{align}
    \sigma_{V_{ab}}^2 &= \left| \frac{\partial{V_{ab}}}{\partial{g_a}} \right|^2 \sigma_{g_a}^2 +
    \frac{\partial{V_{ab}}}{\partial{g_a}}\overline{\frac{\partial{V_{ab}}}{\partial{g_b}}} \text{cov}_{g_bg_a}\nonumber \\
                      &+
                      \overline{\frac{\partial{V_{ab}}}{\partial{g_a}}}\frac{\partial{V_{ab}}}{\partial{g_b}}\text{cov}_{g_ag_b}
        + \left| \frac{\partial{V_{ab}}}{\partial{g_b}} \right|^2 \sigma_{g_b}^2,
    \label{eq:vis_unc_1}
\end{align}

\noindent and since we assume the gains are unity, meaning the coefficients of the variance and
covariance terms are the same, this expression can be simplified further to

\begin{equation}
    \sigma_{V_{ab}}^2 = \left| \frac{\partial{V_{ab}}}{\partial{g_a}} \right|^2 (\sigma_{g_a}^2 + cov_{g_ag_b} +
    cov_{g_bg_a} + \sigma_{g_b}^2).
    \label{eq:simplified}
\end{equation}

The variance and covariance terms can be directly taken from the diagonal and off-diagonals of
the CRB matrix, respectively.

\section{Computational Method}
\label{computational_method}

\subsection{Metrics}
The evaluation of both the image and most of the calibration metrics is simple. Equations
\ref{eq:rms} and \ref{eq:dr} were evaluated after reading in the image data. The smoothness of the
amplitude solutions of the antenna were found by evaluating Equation \ref{eq:smoothness}, after
applying a Blackman-Harris window to the solutions. The phase linearity metrics were evaluated for
each set of phase solutions by first applying a linear fit to the data, then evaluating Equation
\ref{eq:rmse} and Equation \ref{eq:mad}. The Euclidean distance was simply the average of the
absolute value of the difference between the NS and EW phase solutions at each channel, and the KS
metric was obtained using the \texttt{kstest} method from \texttt{scipy} (\cite{Virtanen2020}).

\subsection{Detecting metric outliers}
\label{thresholding}
It is inevitable that some observational data are corrupted by some source of error that is
independent of the sky, such as problems with the telescope or strong RFI. For this work, these
effects need to be disentangled from sky-based effects. Processing these bad observations will
result in outliers in the metrics. These outliers were detected by first treating each observation's
metrics as a 128-dimensional point then calculating the distance to every other observation's metrics
(another 128-dimensional point). We then calculate the MAD for these distances to use in the
modified z-score. In this paper, a threshold of $5\sigma$ was used to determine if a result is an
outlier. The number of observations left after applying this method for each field is given in
Table \ref{tab:obs_left}.

\begin{table}
    \caption{Number of 2-minute observations per field remaining after bad data have been removed
    with the method described in Section \ref{thresholding}.}\label{tab:obs_left}
    \begin{center}
        \begin{tabular}[c]{l|c}
            \hline
            \multicolumn{1}{c|}{\textbf{Field name}} & 
            \multicolumn{1}{c}{\textbf{Num. observations}} \\
            \hline
            EoR0 & 39 \\
            EoR1 & 25 \\
            HERA LST 2.0 & 19 \\
            HERA LST 5.2 & 63 \\
            SKAEOR5 & 65 \\
            SKAEOR6 & 21 \\
            SKAEOR14 & 28 \\
            SKAEOR15 & 68 \\
            \hline
        \end{tabular}
    \end{center}
\end{table}

\subsection{Gain uncertainties}
\label{theoretical_gain_uncertainties}

In this work, all simulations are zenith pointed meaning simulations of some of the fields may not
be representative of reality, as some fields do not transit the zenith of the MWA. However, this
allows us to compare the effects of source positions within the beam between the different fields.
Furthermore, it is common to calibrate only with sources within the main lobe of the primary beam,
because these are the most reliable. Hence, during these calculations, only sources within the main
lobe are preserved. The main lobe will change slightly depending on pointing, but overall it will
behave similarly whether at zenith or off-zenith. Therefore, simulating at zenith will still provide
valuable insight while allowing for much easier implementation.

Additionally, all components in the sky model are treated as point sources in the simulation, as
Equation \ref{eq:final_FIM} requires only the brightness and positions of the sources. This means
that Fornax A, which is composed of shapelets in our sky model, is modelled as a set of point
sources with large brightness. Moreover, systematics like sky model uncertainties are unaccounted
for; all other parameters are assumed to be accurate.

Equation \ref{eq:radiometer_full} was first evaluated for the given input parameters. In this work the
values used are representative of EoR science with the MWA telescope $T_\text{sys} =
\SI{200}{\kelvin}$ (value at the centre of the bandwidth), $N = 128$, $A_\text{eff} =
\SI{21}{\m\squared}$, $\Delta\nu = \SI{80}{\kilo\hertz}$, and $\Delta\tau = \SI{8}{\second}$. The
field of view of a telescope can be approximated with $\theta \approx \frac{\lambda}{D}$ where $D$
is the diameter of a dish/antenna. For the MWA, $D = \SI{4.4}{\meter}$. We have assumed that
$A_{\text{eff}}$ does not change with frequency (\cite{Tingay2013}).

Next, the calculation and propagation of the gain uncertainties takes place within a large loop over
the frequency range. At the beginning of each iteration, sources in the sky model were vetoed by the
current field of view and also their estimated flux density for that frequency. Following this, the
beam pattern is calculated for the current frequency, using the array factor method
(\cite{Warnick2018}), in a 1024 by 1024 grid where each pixel corresponds to a portion of the sky in
the $(l, m)$ plane. These pixel values are used to attenuate sources located within the pixel by
simply multiplying the source brightness by the pixel value.

The attenuated sources are then used to calculate the CRB matrix by first generating the FIM matrix
with Equation \ref{eq:final_FIM}, where each element represents an antenna pair $(a, b)$. The FIM is
Hermitian, hence only the top triangle of the matrix needs to be calculated and the bottom triangle
can be filled in by taking the complex conjugate. The CRB are the diagonal elements of the inverse
of the FIM matrix, and the gain uncertainties for each antenna are their square-root.

\vspace{2em}

\subsection{Power spectra}
Taking the previously calculated gain uncertainties, still within the large frequency loop, the
uncertainties for visibilities were evaluated with Equation \ref{eq:simplified}. Once uncertainties were
calculated, the real and complex components for the visibilities for each baseline were randomly
generated from a normal distribution with mean 0 and standard deviation
$\frac{\sigma_{V_{ab}}}{\sqrt{2}}$. The visibilities were then gridded onto a common $(u,v)$ grid, and
appended onto a growing list of gridded visibilities. 

Once the loop completed, we were left with a data cube of gridded visibilities at each frequency
channel. To transform this data cube into the final power spectrum, a Fourier Transform was applied
to each $(u,v)$ cell along the frequency axis. The result was then multiplied by its complex
conjugate to yield the unnormalised power. We then cylindrically average the unnormalised power at
each $\eta$ slice (the FT of frequency) to arrive at the 2D temperature power spectrum.


\section{Results}
\label{results}

This section will present the results of the calculations and attempt to provide explanations for the
behaviour seen in some metrics. The information gained and the usefulness of the metrics will also
be discussed. All results shown are generated after applying the threshold technique discussed in
Section \ref{thresholding}. Outliers in this study generally stem from either computational errors
or bad data, not from the behaviour of the fields. Differences between NS and EW polarisations arise
from their different fringe patterns. {In this section, we display results only for
the EoR0 field as an indicative field; results for the other fields can be found in the appendices}.
The following metrics were found to be most useful: image RMS and dynamic range, amplitude
smoothness, and phase RMSE.
\subsection{Image metrics}

\begin{figure}[h]
    \begin{center}
        \includegraphics[width=0.95\textwidth]{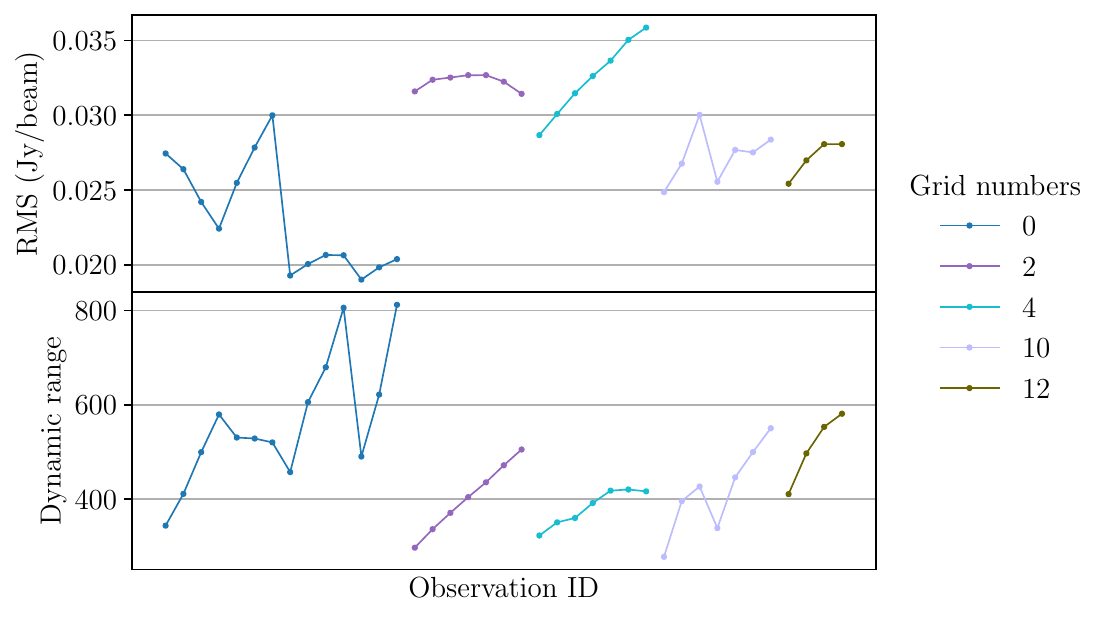}
    \end{center}
    \caption{Image metrics for the EoR0 field grouped by pointing (represented by different colours)
    for observation IDs in ascending order. The top panel shows the Root Mean Square (RMS) metric.
    The bottom panel shows the Dynamic Range metric. The changing values between observations are
    due to sources moving in and out of the MWA beam.
metric.}\label{fig:EoR0_image_metrics}
\end{figure}

The RMS and dynamic range of the EoR0 field can be found in Figure \ref{fig:EoR0_image_metrics}. Both the
RMS and DR of the fields can be seen varying with the observation ID for a particular pointing. This
behaviour could be due to sources moving in and out of the main/side lobes of the MWA with
observation ID. Although these metrics are displayed in increasing observation ID, it does not
necessarily mean successive points are observations performed immediately after each other.

\subsection{Amplitude solutions smoothness}

\begin{figure}[h]
    \begin{center}
        \includegraphics[width=0.95\textwidth]{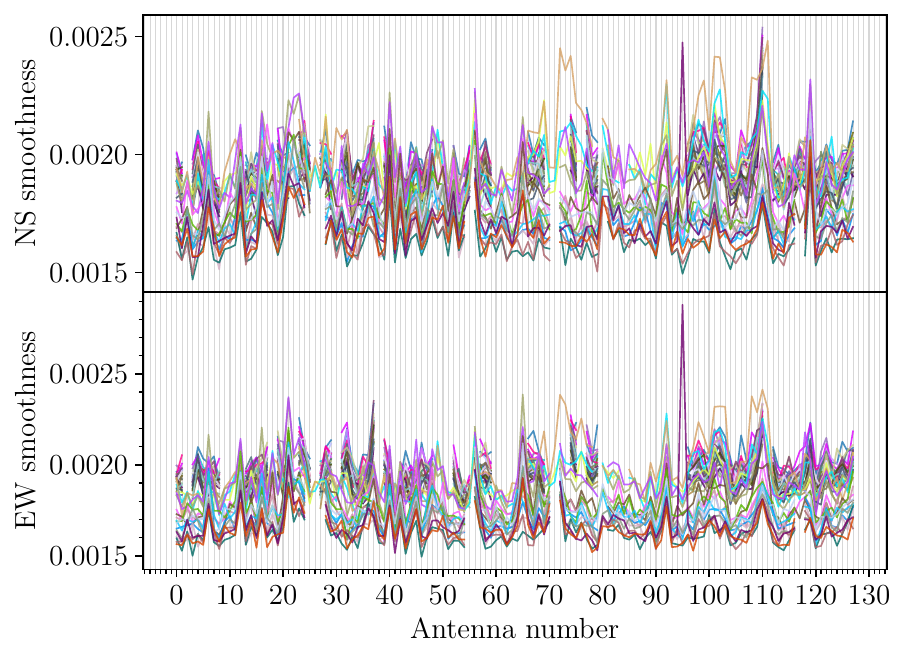}
    \end{center}
    \caption{Smoothness of the NS (top) and EW (bottom) calibration amplitude solutions for
    EoR0 for each antenna. Each line represents a different observation ID. A smaller value
represents smoother amplitude solutions.}\label{fig:EoR0_xx_smoothness}
\end{figure}

The smoothness of both NS and EW polarisation amplitude calibration solutions for EoR0 can be found
in Figure \ref{fig:EoR0_xx_smoothness}. Some fields exhibit clustering between lines of different
observations IDs, this is especially prominent in the HERA LST 2.0 field in the NS amplitude
smoothness found in \ref{amplitude_smoothness}. This clustering behaviour could be explained by
different conditions at the time of observation or, simply, an insufficient amount data to provide a
robust representation of the field.

\begin{figure}[h]
    \begin{center}
        \includegraphics[width=0.95\textwidth]{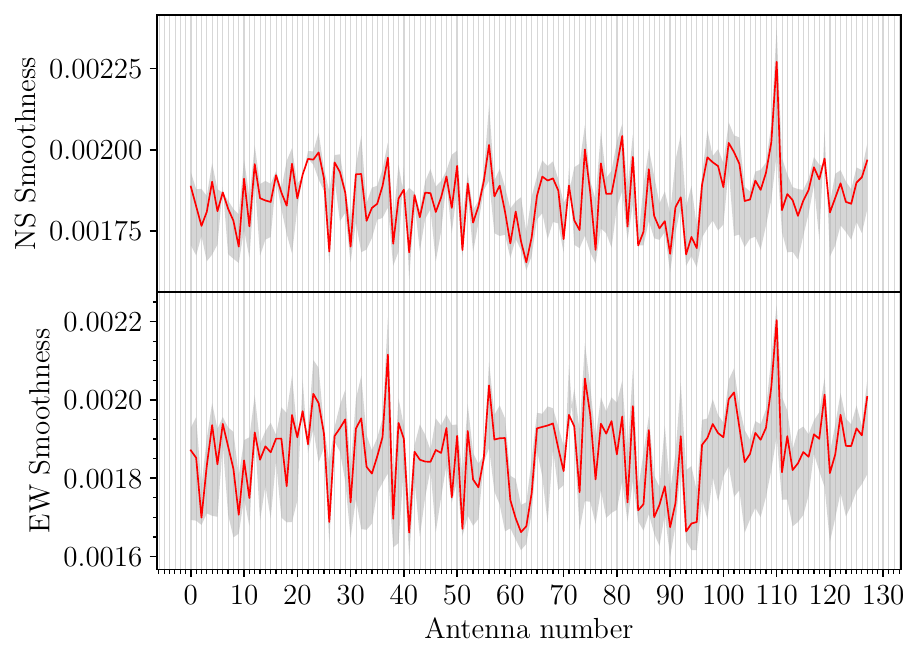}
    \end{center}
    \caption{Band plot of the smoothness of the NS (top) and EW (bottom) amplitude calibration
    solutions for EoR0 for each antenna. The median per antenna is given in red, the upper and lower
quartiles per antenna are given by the shaded region. A smaller value represents smoother amplitude
solutions.}\label{fig:EoR0_xx_smoothness_band}
\end{figure}

For a simplified version of the same data, a band plot of the same EoR0 data in
Figure \ref{fig:EoR0_xx_smoothness} is given in Figure \ref{fig:EoR0_xx_smoothness_band}, plotted in this figure
are the median (red line), and the lower and upper quartiles (shaded region).

\begin{figure}
    \begin{center}
        \includegraphics[width=0.95\textwidth]{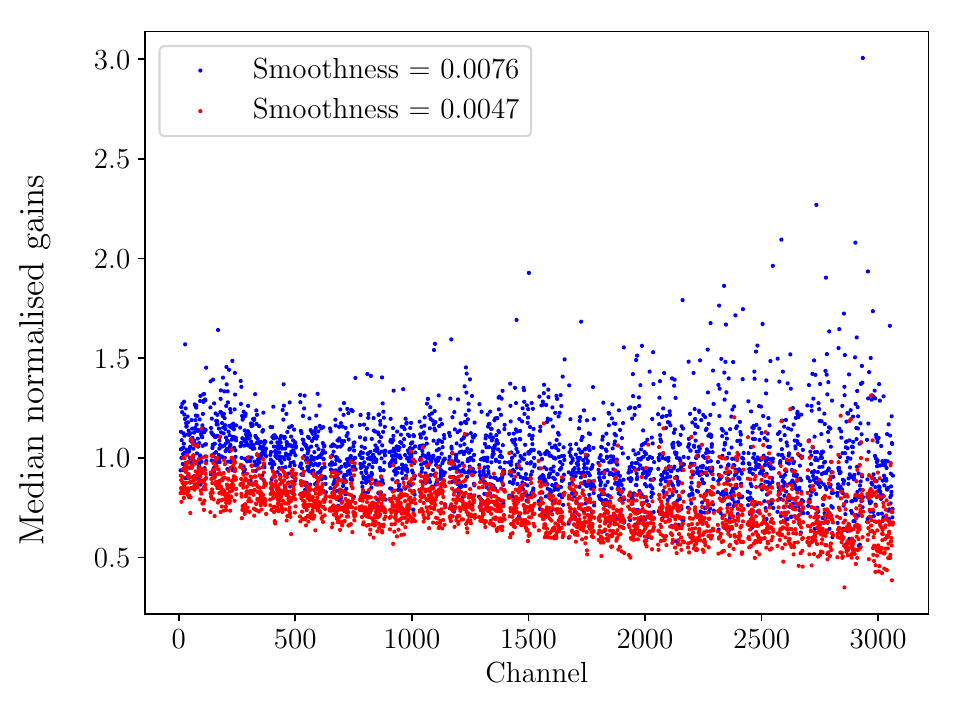}
    \end{center}
    \caption{Calibration amplitude solutions for antenna 100 (red points) and antenna 110 (blue
    points) of observation 1201153128. A lower smoothness metric corresponds to visually smoother
    amplitudes. Indeed, we see that the antenna 100 ($\text{smoothness} = 0.0047$) is visually
smoother then antenna 110 ($\text{smoothness} = 0.0076$). }
    \label{fig:good_bad_antenna}
\end{figure}

To showcase the smoothness metric, the NS amplitude solutions are given for observation
$1201153128$ at antennas $100$ and $110$ in Figure \ref{fig:good_bad_antenna}. From
Figure \ref{fig:EoR0_xx_smoothness_band} we can see that, generally, antenna 110 amplitude solutions are less
smooth than the solutions of antenna 100. Indeed, this is the observed behaviour in
Figure \ref{fig:good_bad_antenna}.

\subsection{Phase solutions metrics}

\begin{figure}[h]
    \begin{center}
        \includegraphics[width=0.95\textwidth]{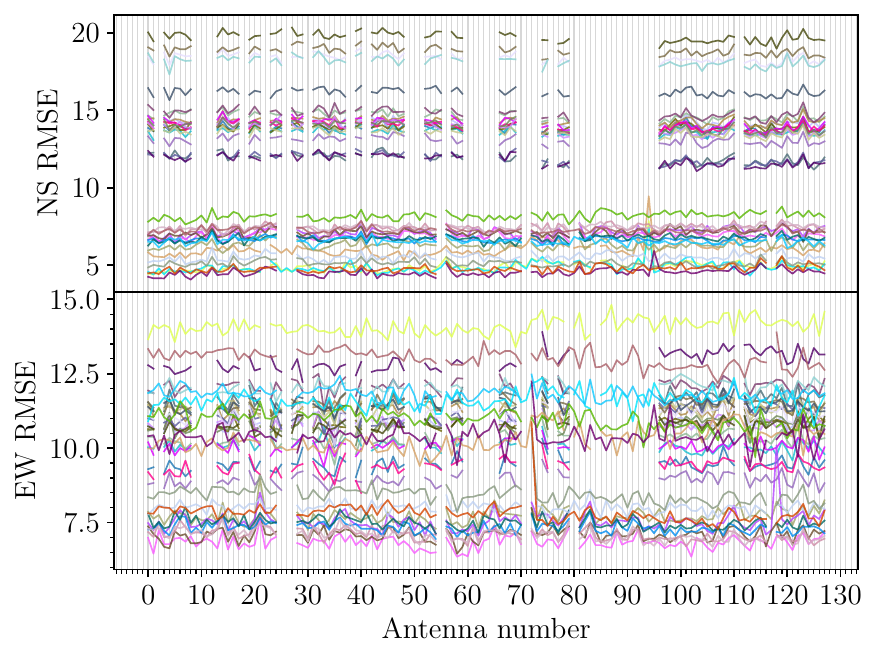}
    \end{center}
    \caption{Root-Mean-Square-Error metric of both NS (top) and EW (bottom) calibration phase
        solutions for observations of the EoR0 field. Each line represents a different observation
    ID. A smaller value represents a more linear phase solution.}\label{fig:EoR0_rmse}
\end{figure}

The RMSE metrics for EoR0 are displayed in Figure \ref{fig:EoR0_rmse}. In this metric, a smaller number
corresponds to a more linear phase solution. We once again observe similar clustering behaviour in
some of these results. However, it appears that behaviour that is present in the amplitude
calibration solutions does not necessarily translate to the phase metrics. The MAD behaves almost
identically to the RMSE, hence, moving forward we focus on the RMSE metric.

\begin{figure}[h]
    \begin{center}
        \includegraphics[width=0.95\textwidth]{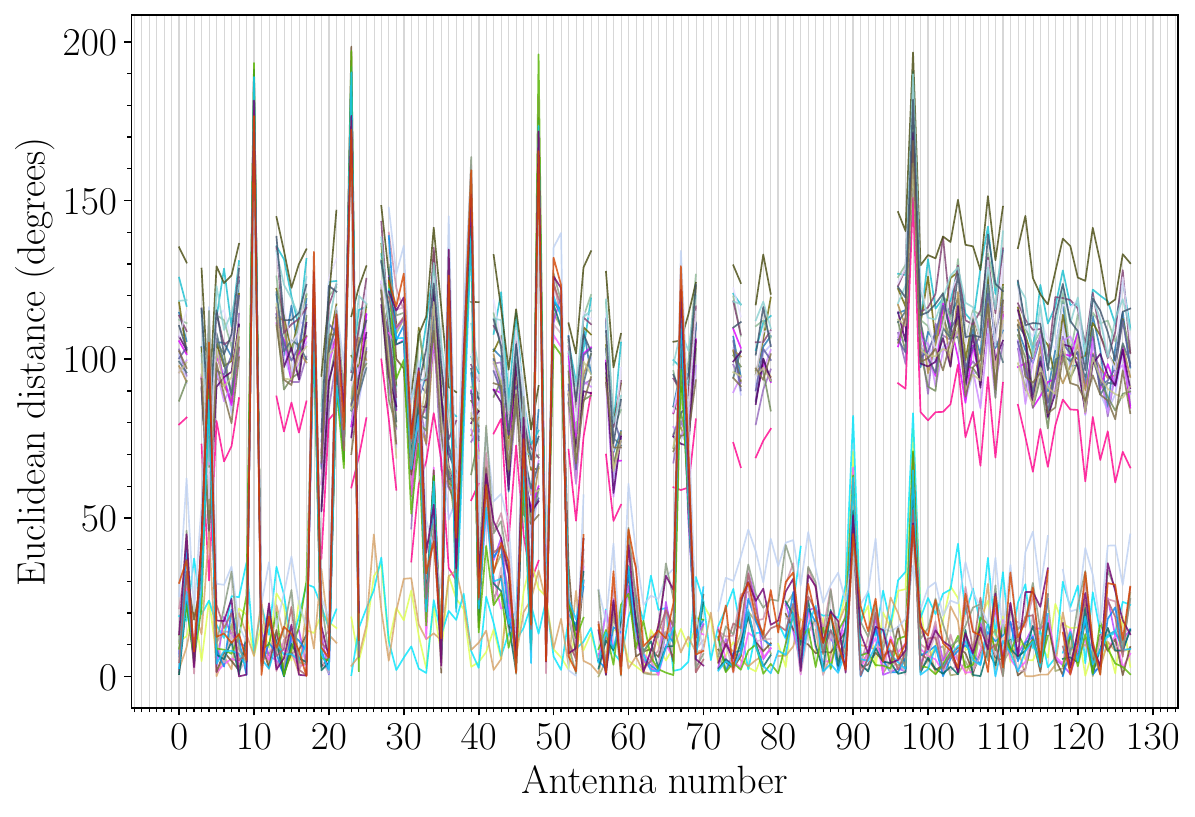}
    \end{center}
    \caption{The average Euclidean distance metric between NS and EW calibration phase solutions for
    the EoR0 field. Each line represents a different observation ID. A lower value represents more
similar phase solutions. In this field, there appears to be two groupings of distances where one
group describes very similar phase solutions.}\label{fig:EoR0_euclid}
\end{figure}

Finally, the average Euclidean distance metrics for the EoR0 field are given in
Figure \ref{fig:EoR0_euclid}. It is challenging to glean any information from the two similarity metrics,
in particular the KS-metric seems to be the most random of all the metrics. For that reason, we will
continue without KS-metric. However, in the average Euclidean distance metric some fields do exhibit
clustering behaviour. For example, the EoR0 field appears to have two distinct groupings of
average-distances, where one group is more similar than the other. Although it does exhibit more
behaviour than the KS-metric, we opt to move without this metric going forward, as it is still
difficult to obtain useful information.

\vspace{1em}

\subsection{Smoothness and RMSE correlations}
\begin{figure}
    \begin{center}
        \includegraphics[width=0.95\textwidth]{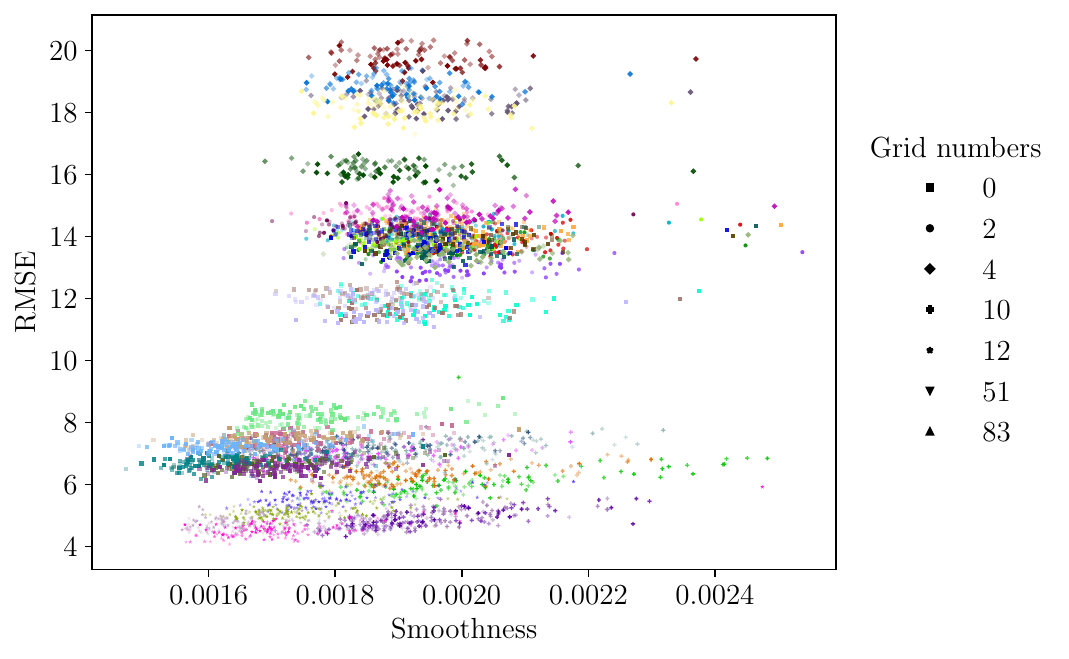}
    \end{center}
    \caption{Plot displaying the correlation between the calibration amplitude smoothness metric and
        the NS calibration phase Root-Mean-Square-Error (RMSE) metric for observations of the EoR0
        field. There is a positive correlation between the two metrics and seems to flatten off at
        higher RMSE values. Also seen is clustering of the observations with grid number 0, at RMSE
        values of 7, 12, and 14.}\label{fig:EoR0_xx_smoothness_rmse}
\end{figure}

The NS smoothness and RMSE correlation plot for the EoR0 field can be found in
Figure \ref{fig:EoR0_xx_smoothness_rmse}. In this figure, each colour and marker combination indicates an
individual observation. Each point is described by the antenna's smoothness and RMSE metrics, and
the opacity of the point relates to the antenna number. The more opaque points correspond to the
long baseline tiles. In this figure, we can see for observations below RMSE=8, there exists a
correlation between the smoothness and RMSE metric. As the amplitude solutions become less smooth
(increasing smoothness metric) the RMSE metric also increases. This trend seems to flatten off for
observations above RMSE=8 in the EoR0 field. Results for other fields, although not exactly the same
shape, all show there is a positive correlation between these two metrics.

It is also worth noting in Figure \ref{fig:EoR0_xx_smoothness_rmse} the clustering of
observations with grid number 0 around RMSE values of 7, 12, and 14. A table of LSTs and start dates
in UTC for observations with grid number 0 are given in Table \ref{tab:EoR0_LST}. The cause of this
clustering behaviour seem to stem from conditions at the time of observation and how close the
observations are in terms of time.

\begin{table}
    \caption{Table of EoR0 observations with a grid number 0. The LST and start date in UTC are
    included for each observation. There are at least two large groups of observations which are
close in time, while the others form smaller groups. These groupings could provide an explanation
for the clustering of observations seen in Figure \ref{fig:EoR0_xx_smoothness_rmse}.
}\label{tab:EoR0_LST}
    \begin{center}
        \begin{tabular}[c]{l|c|c}
            \hline
            \multicolumn{1}{c|}{\textbf{Observation ID}} & 
            \multicolumn{1}{c}{\textbf{LST}} &
            \multicolumn{1}{c}{\textbf{Start date (UTC)}} \\
            \hline
            1286204984 & 361.98 & 2024-09-24T10:47:08 \\
            1286203784 & 356.97 & 2024-09-24T10:48:58\\
            1286205584 & 364.49 & 2024-09-24T10:46:57\\
            1286806936 & 356.98 & 2024-09-24T10:46:54\\
            1286807536 & 359.49 & 2024-09-24T09:43:57\\
            1194267224 & 358.92 & 2024-09-24T13:09:09\\
            1194266984 & 357.91 & 2024-09-24T13:12:29\\
            1286808736 & 364.50 & 2024-09-24T09:29:54\\
            1194267464 & 359.92 & 2024-09-24T13:05:15\\
            1286204384 & 359.48 & 2024-09-24T10:48:50\\
            1194267704 & 0.92 & 2024-09-24T13:02:45\\
            1194266744 & 356.91 & 2024-09-24T13:15:29\\
            1194267944 & 1.92 & 2024-09-24T12:00:17\\
            1194268184 & 2.93 & 2024-09-24T11:54:31\\
            \hline
        \end{tabular}
    \end{center}
\end{table}

\subsection{Theoretical gain uncertainties}

\begin{figure}
    \begin{center}
        \includegraphics[width=0.95\textwidth]{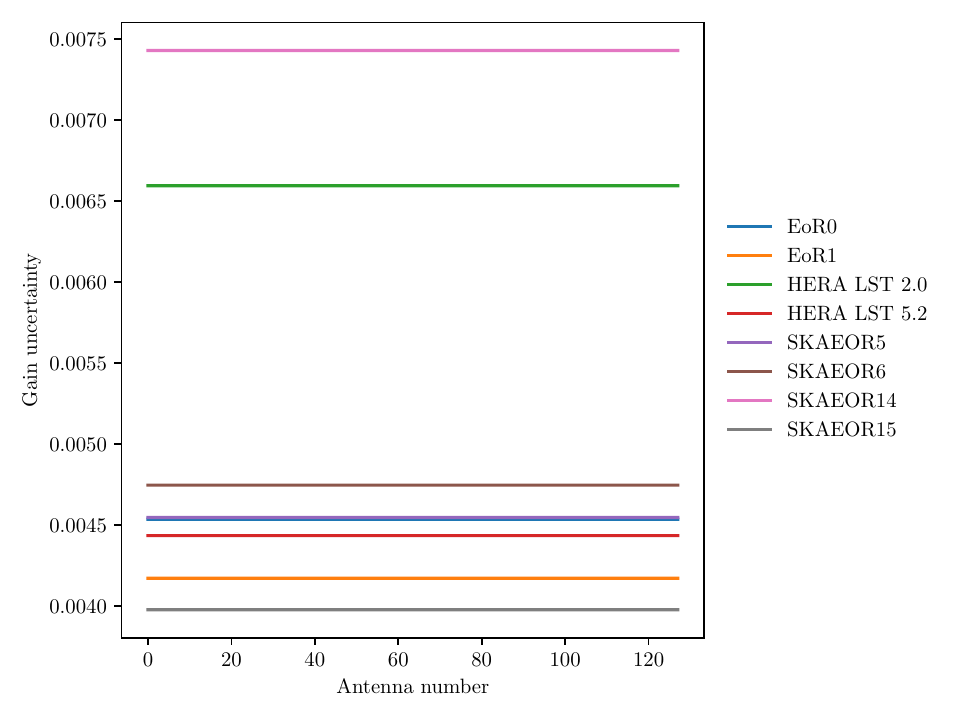}
    \end{center}
    \caption{Theoretical gain uncertainties of various fields for 128 MWA antennas in the phase II
configuration at $\SI{182}{\mega\hertz}$, calculated with the procedure described in
Section \ref{theoretical_gain_uncertainties}. Each field were treated as being zenith pointed. There are
miniscule fluctuations between antennas for all fields.}\label{fig:gain_uncertainties} \end{figure}

\begin{figure}
    \begin{center}
        \includegraphics[width=0.95\textwidth]{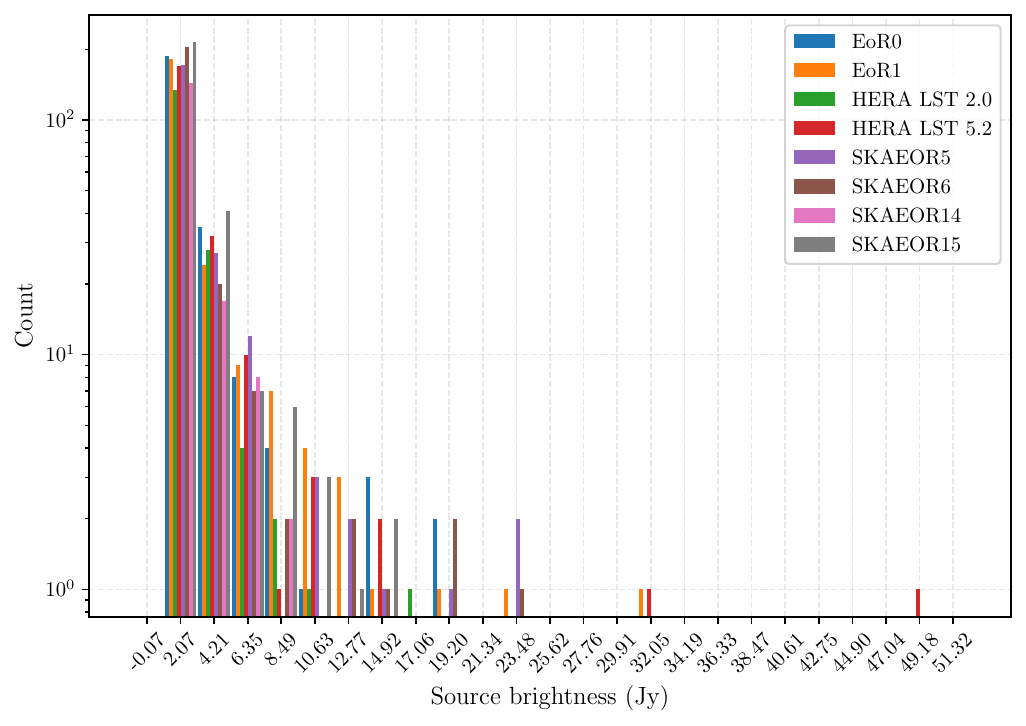}
    \end{center}
    \caption{Combined histogram of source brightness for each field at $\SI{182}{\mega\hertz}$
    during the Crame\'r-Rao Bounds calculation. The width of each bin is $2.14\, \text{Jy}$. This
histogram, along with Table \ref{tab:field_stats_table}, help to investigate how the distribution of
source brightness result in the gain uncertainties seen in Figure \ref{fig:gain_uncertainties}.}
\label{fig:combined_histogram} \end{figure}

The theoretical gain uncertainties for each field as a function of antenna number can be found in
Figure \ref{fig:gain_uncertainties}. A combined histogram of the source brightness in each field is given
in Figure \ref{fig:combined_histogram} and the number of sources in each field along with the brightest
source is given in Table \ref{tab:field_stats_table}. The three fields that produce the best gain
uncertainties via the CRB matrix are SKAEOR15, EoR1, and HERA LST 5.2. These three fields show that
having bright sources and a large number of points increase the gain precision.

\begin{table} \caption{Table of the number of sources within the field of view of the beam and the
    brightest source at $\SI{182}{\mega\hertz}$ during the Crame\'r-Rao Bound calculation, for each
field. This table helps us reveal how the number of sources, and the brightest source result in
the gain uncertainties we see in Figure \ref{fig:gain_uncertainties}.}\label{tab:field_stats_table}
\begin{center}
        \begin{tabular}[c]{l|c|c}
            \hline
            \multicolumn{1}{c|}{\textbf{Field name}} & 
            \multicolumn{1}{c}{\textbf{Number of sources}} &
            \multicolumn{1}{c}{\textbf{Brightest source (Jy)}} \\
            \hline
            EoR0 & 241 & 19.3 \\
            EoR1 & 233 & 32.4 \\
            HERA LST 2.0 & 170 & 16.2 \\
            HERA LST 5.2 & 220 & 50.0 \\
            SKAEOR5 & 219 & 23.3 \\
            SKAEOR6 & 240 & 23.3 \\
            SKAEOR14 & 170 & 8.6 \\
            SKAEOR15 & 274 & 14.2 \\
            \hline
        \end{tabular}
    \end{center}
\end{table}

\begin{figure}
    \begin{center}
        \includegraphics[width=0.95\textwidth]{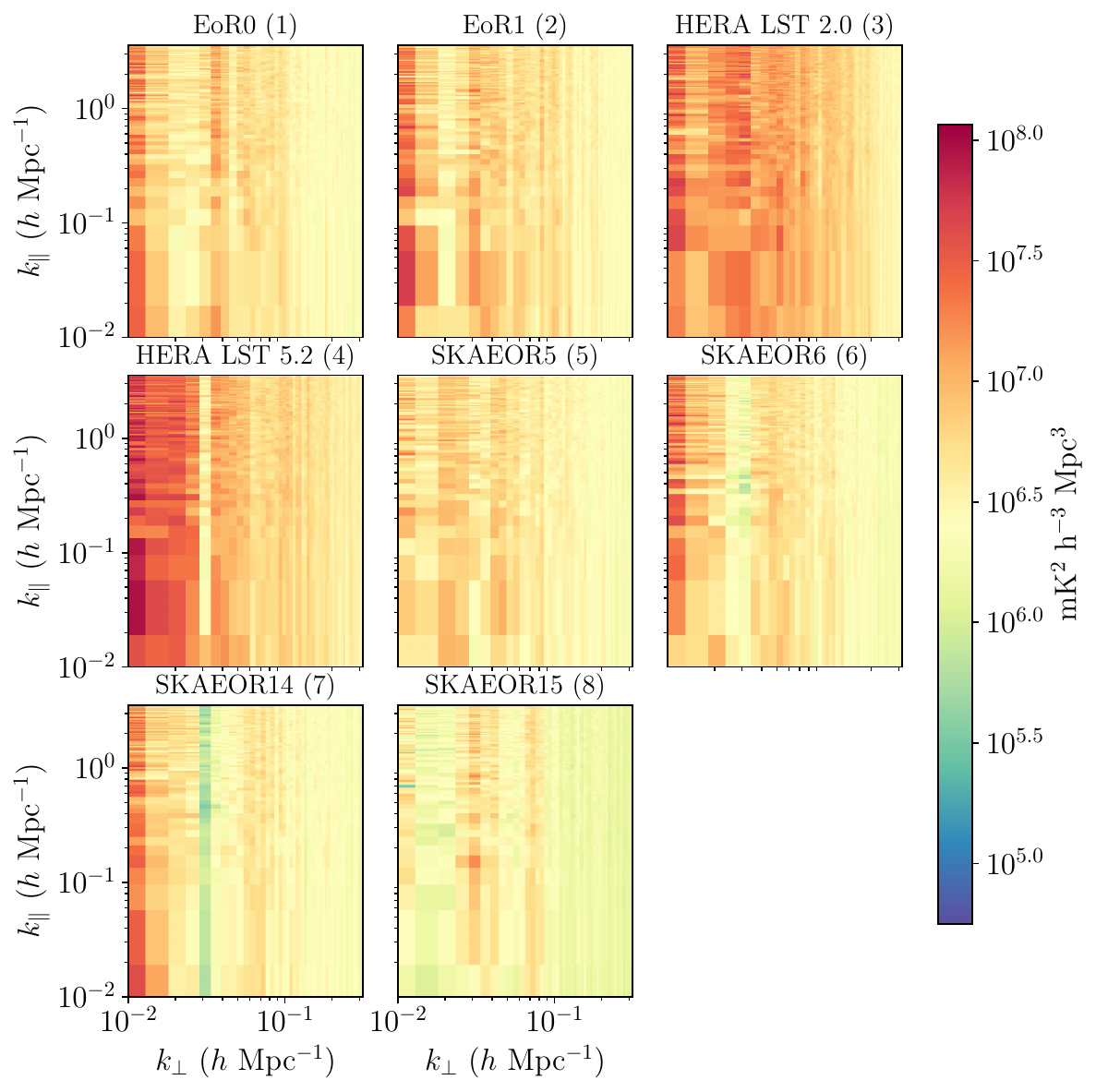}
    \end{center}
    \caption{Resulting power spectra for each field after propagating theoretical uncertainties into
    visibilities. Each field was treated as if it were at zenith.}\label{fig:multi_power_spec}
\end{figure}

The power spectra for each field, after propagating these theoretical uncertainties into the
visibilities, can be found in Figure \ref{fig:multi_power_spec}. Interestingly, fields which have shown
better theoretical calibration solutions previously do not have the best performing power spectra.
This can be explained by considering Equation \ref{eq:simplified}, the partial derivative coefficient is
dependent on the brightness of the sources selected by the simulation. Hence, even if the gain
uncertainties may be better for a particular field, they are imprinted upon the brightest sources in
the field which may mean the power spectrum overall performs worse in comparison to other fields.

\begin{figure}
    \begin{center}
        \includegraphics[width=0.95\textwidth]{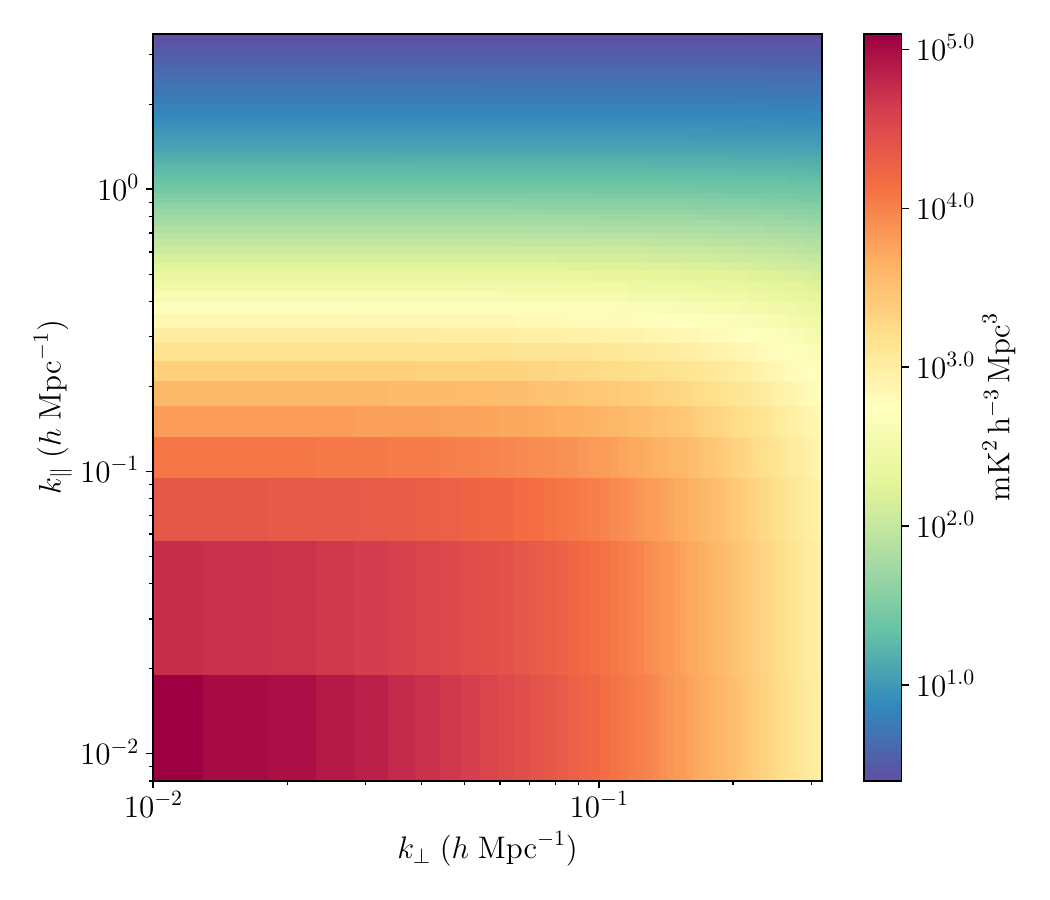}
    \end{center}
    \caption{A 2D power spectrum of a pure \SI{21}{\centi\meter} signal from a `faint
    galaxies' simulation (\cite{Mesinger2016}) centred on $z \sim 7$.} \label{fig:deprojected}
\end{figure}

A 2D power spectrum for a pure \SI{21}{\centi\meter} signal from a `faint galaxies' model
(de-projected from a 1D spherically averaged power spectrum, \cite{Mesinger2016}) is given in Figure
\ref{fig:deprojected}. The theoretical power spectra of this work are at least an order of magnitude
larger than the predicted signal.

\section{Discussion}
\label{discussion}

Having displayed the metrics and the theoretical analysis, we will attempt to
select a few fields that may be ideal for EoR science. We will first begin by disentangling
telescope-based effects from sky-based effects. Hence, our first criteria will be the number of
observations available per field. This leaves us with four fields, these fields are: EoR0, HERA LST
5.2, SKAEOR5, and SKAEOR15. Within these fields, we inspect the average behaviour of the metrics,
not the antenna to antenna or observation to observation behaviour, which should be uncorrelated.
Figures for image metrics of the other metrics can be found in \ref{image_metrics}, amplitude
smoothness in \ref{amplitude_smoothness} and \ref{smoothness_bands}, phase RMSE in
\ref{rmse_metrics}, and average Euclidean distance in \ref{euclid}.

Next, we narrow down the selection based on the performance of these four fields in the metrics.
Table \ref{tab:performance} displays how the fields compare relative to each other, where the
comparison of metrics was a simple visual inspection. From this table, we can immediately exclude
SKAEOR5 and SKAEOR15 from our selection, leaving only the EoR0 and HERA LST 5.2 fields.

\begin{table*}[h]
    \caption{Table of the four fields for consideration as an EoR observing field, and their
        relative performance to each other in the metrics used in this work. Performance was
        determined visually based on desired behaviours as described in Table \ref{tab:metrics}. A
        check mark ($\checkmark$) indicates the respective field(s) perform well in that metric, a
        tilde ($\sim$) indicates reasonable performance, and a cross ($\times$) indicates bad
        performance.}
    \label{tab:performance}
    \begin{tabular}[c]{c|c|c|c|c|c|c}
         & \multicolumn{6}{c}{\textbf{Metrics}} \\ \cline{2-7}
        \textbf{Field} & Image RMS & Image DR & XX Amp. Smoothness & YY Amp. Smoothness & XX Phase
        RMSE & YY Phase RMSE \\ \hline
        EoR0 & $\checkmark$ & $\checkmark$ & $\sim$ & $\sim$ & $\times$ & $\checkmark$ \\ \hline
        HERA LST 5.2 & $\times$ & $\checkmark$ & $\checkmark$ & $\checkmark$ & $\checkmark$ & $\times$ \\ \hline
        SKAEOR5 & $\sim$ & $\sim$ & $\times$ & $\times$ & $\times$ & $\times$ \\ \hline
        SKAEOR15 & $\times$ & $\times$ & $\times$ & $\times$ & $\sim$ & $\checkmark$ \\ \hline
    \end{tabular}
\end{table*}

Introducing the power spectra now will further narrow down the selection. To serve as a reminder,
these power spectra showcase the impact of gain uncertainties due to the calibrators' positions
within the beam. They also simulate a zenith pointing, hence simulations of these fields are not
completely physical. Additionally, due to the sky model used in this study, some parts of the sky
are better modelled than others.

With these factors in mind, from Figure \ref{fig:multi_power_spec} it is clear that the HERA LST 5.2 field
is most affected by the gain uncertainties. The other three fields: EoR0, SKAEOR5, and SKAEOR15,
display acceptable impacts from gain uncertainties. However, in reality, the SKAEOR5 field has a very
northern declination, where the beam of the MWA starts to misbehave. Taking everything into account,
the EoR0 field seems to be a safe field for observation.

We must also discuss the applicability of these metrics, which were calculated using archived MWA
data, to the SKA-Low telescope. SKA-Low will have a much smaller field-of-view, $\sim\SI{2.5}{\degree}$ as
opposed to $\sim\SI{21}{\degree}$ at $\SI{184.32}{\mega\hertz}$, along with better
sensitivity than the MWA. To compare the number of sources that would be detected by the MWA and
SKA-Low, we can investigate the ratio of the integrated source counts,
$\frac{N_{\text{SKA}}}{N_\text{MWA}}$. The integrated source count is given by

\begin{equation}
    N(S>S_\text{min}) = \frac{\alpha}{1 - \beta} \left( \frac{S_\text{min}}{S_0} \right)^{1-\beta}
    \Omega, 
    \label{eq:source_count}
\end{equation}

\noindent where $\alpha$ and $\beta$ are fitting parameters that are intrinsic to the sky. The FOV,
$\Omega$, can be approximated by $\frac{\lambda}{D}$, where $D$ is the diameter of an MWA tile or
SKA station. $S_{\text{min}}$ is the sensitivity of the MWA or SKA is given by Equation
\ref{eq:radiometer_full}. In this section we express the sensitivity in the form

\begin{equation}
    S_{\text{min}} = \frac{\text{SEFD}}{\sqrt{\Delta \nu \Delta \tau} \sqrt{N_b}},
    \label{eq:sensitivity}
\end{equation}

\noindent where $\Delta \nu$, $\Delta \tau$, and $N_b$ are the bandwidth, integration times, and
number of baselines respectively. The SEFD is calculated using 

\begin{equation}
    \text{SEFD} = \frac{2 k T_{\text{sys}}}{A_e},
    \label{eq:SEFD}
\end{equation}

\noindent where $k$ is the Boltzmann constant, and $A_e$ is the effective collecting area. The ratio
of SKA integrated source counts to MWA integrated source counts is given by


\begin{equation}
    \frac{N_{\text{SKA}}}{N_{\text{MWA}}}
    =\left(\frac{\text{SEFD}_\text{SKA}\sqrt{N_{b,\text{MWA}}}}{\text{SEFD}_\text{MWA}\sqrt{N_{b,\text{SKA}}}}\right)^{1-\beta}
    \frac{D_\text{MWA}}{D_\text{SKA}},
    \label{eq:ratio_source_counts}
\end{equation}

\noindent where $D_{\text{SKA}}$ and $D_{\text{MWA}}$ are the diameters of an SKA station and MWA
tile respectively (approximately 38m for the SKA and 4.4m for the MWA). For both SKA and MWA we
assume $T_{\text{sys}} \approx \SI{200}{\kelvin}$ at $\SI{150}{\mega\hertz}$. The effective
collecting area of the MWA at $\SI{150}{\mega\hertz}$ is $A_{e,\text{MWA}} =
\SI{21.5}{\metre\squared}$, for the SKA we will assume a single SKA station is 100\% efficient
resulting in $A_{e,\text{SKA}} = \SI{1134.11}{\metre\squared}$. The system equivalent flux densities
are then $\text{SEFD}_{\text{MWA}} \approx 25000 \,\text{Jy}$ and $\text{SEFD}_{\text{SKA}} \approx
500 \,\text{Jy}$. In this calculation we use the source counts of \cite{Intema2011} to obtain $\beta
= 1.59$. The number of unique baselines for SKA-Low and MWA are $N_{b,\text{SKA}} = 131086$ and
$N_{b,\text{MWA}} = 8128$. With these values, Equation \ref{eq:ratio_source_counts} is evaluated to
approximately $\frac{N_{\text{SKA}}}{N_{\text{MWA}}} \approx 0.5$. 

The SKA-Low will approximately detect half of what the MWA can detect, with the better sensitivity
and smaller field of view. The effects of this will be difficult to discuss with
certainty. But since we have shown a dependence on the number of sources and in particular the
brightest sources, fields with bright sources near zenith in both the SKA and MWA field of view can
be expected to behave similarly. Additionally, there is a dependence on the number of antennas in
Equation \ref{eq:final_FIM}. The combined effects of these dependencies is difficult to predict.

In the image metrics with SKA-Low, we can expect lower RMS and larger dynamic range (if a bright
source is within the field of view) due to the increased sensitivity of SKA-Low. The calibration
metrics are more challenging to estimate. We have seen with the HERA LST 5.2 simulation
(which, in reality, is a zenith field) that the bright source contributes to a lower uncertainty,
which is then reflected in the NS and EW amplitude smoothness metrics. Considering the correlation
between amplitude smoothness and phase RMSE, we can further reason that the phase solutions should
also be linear. Of course, this line of reasoning is only applied for the HERA LST 5.2 field and may
not hold for other fields. Recently, work from LOFAR which uses a 6-hour observation of their target
field containing 3C196 (\cite{Ceccotti2025}), a very bright source near zenith, has seen lower
systematics compared to the colder NCP field. This aligns with the behaviour we have explored in
this work. We do not apply additional steps in the power spectrum estimation, hence, we cannot
compare the power spectra of this work with those of the recent LOFAR work.

SKA-Low will also have the ability to form multiple beams and utilise sub-stations. It will also use
digital beam formers which will allow for greater pointing precision. Further investigation into how
these variables will affect EoR observations will be needed.

\section{Conclusion}
\label{conclusion}

In this study we have investigated image metrics (RMS and dynamic range) and calibration metrics
(amplitude smoothness, RMSE of the phase solutions, and average Euclidean distance between the phase
solutions of different polarisations). We utilised archival MWA data in the phase II configuration
centred on $\SI{184.32}{\mega\hertz}$ ($z \sim 6.8$), with observations pointing towards fields used
by the MWA collaboration (\cite{Lynch2021}), HERA team (\cite{Abdurashidova2022}), and fields
previously chosen by other metrics (\cite{Zheng2020}). These data were used as a proxy for future
SKA-Low data. We found the most useful metrics to be the image RMS, dynamic range, amplitude
smoothness, and phase RMSE.

In addition to the metrics, a theoretical method utilising the Cr\'amer-Rao bound was used to
calculate theoretical gain uncertainties. A standard error propagation of these uncertainties into
the final power spectrum was also provided. From this we have seen that the brightest sources
contribute the most to a lower gain uncertainty.

The combination of both metrics and theoretical power spectra helped to confirm that EoR0 is indeed
a candidate for future EoR observations. {In particular, the field's performance in the both
image metrics, and its impacts on the final power spectrum --- while also being a real pointing used
by the MWA --- has led us to this conclusion.} However, this shouldn't dissuade the investigation of
the other fields. The HERA LST 5.2 field was shown to perform well in the smoothness metric, which
aligns with behaviour in recent work from LOFAR (\cite{Ceccotti2025}). Additionally, the HERA LST
5.2 field has the largest dynamic ranges (but also large RMS) out of the fields discussed. The
SKAEOR5 field, which contains the second-largest number of observations has a slowly varying image
RMS second to the EoR0 RMS. Clearly, these fields are still worth investigating.

\begin{acknowledgement}
This research was partly supported by the Australian Research Council Centre of Excellence for
All Sky Astrophysics in 3 Dimensions (ASTRO 3D), through project number CE170100013. The
International Centre for Radio Astronomy Research (ICRAR) is a Joint Venture of Curtin
University and The University of Western Australia, funded by the Western Australian State
government. This scientific work uses data obtained from \textit{Inyarrimanha Ilgari Bundara} /
the Murchison Radio-astronomy Observatory. We acknowledge the Wajarri Yamaji People as the
Traditional Owners and native title holders of the Observatory site. Establishment of CSIRO’s
Murchison Radio-astronomy Observatory is an initiative of the Australian Government, with
support from the Government of Western Australia and the Science and Industry Endowment Fund.
Support for the operation of the MWA is provided by the Australian Government (NCRIS), under a
contract to Curtin University administered by Astronomy Australia Limited. This work was
supported by resources provided by the Pawsey Supercomputing Research Centre with funding from
the Australian Government and the Government of Western Australia.
\end{acknowledgement}

\pagebreak

\printendnotes

\bibliography{main.bib}

\newpage
\appendix

\section{Image metrics}
\label{image_metrics}

\begin{figure}[H]
    \begin{center}
        \includegraphics[width=0.95\textwidth]{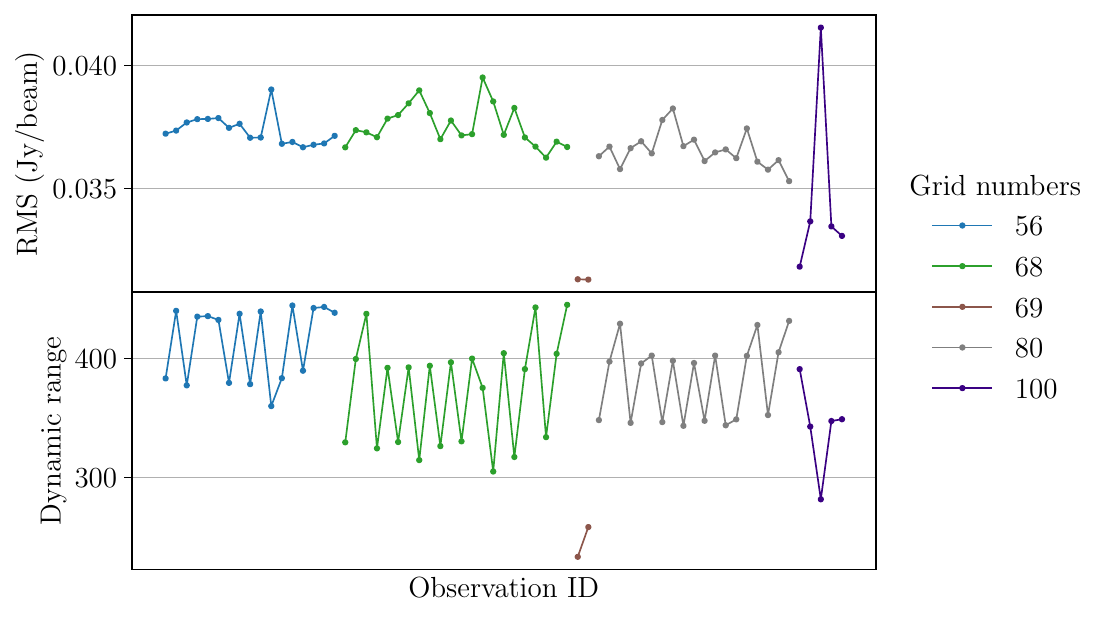}
    \end{center}
    \caption{The RMS of the SKAEOR5 field is shown in the top plot, the dynamic range is shown in the
    bottom plot. Different colours correspond to different grid numbers in the field. Each point
within each pointing corresponds to an observation ID, with observation IDs increasing within a
pointing.}\label{fig:SKAEOR5_image_metrics}
\end{figure}

This appendix contains figures for the image RMS and dynamic range metrics for the HERA LST 5.2,
SKAEOR5, and SKAEOR15 fields for comparison. A low RMS and large dynamic range are ideal behaviours
for these metrics. Interestingly, the SKAEOR5 field (Figure \ref{fig:SKAEOR5_image_metrics})
displays slowly varying RMS over a large number of observations.

\begin{figure}[H]
    \begin{center}
        \includegraphics[width=0.95\textwidth]{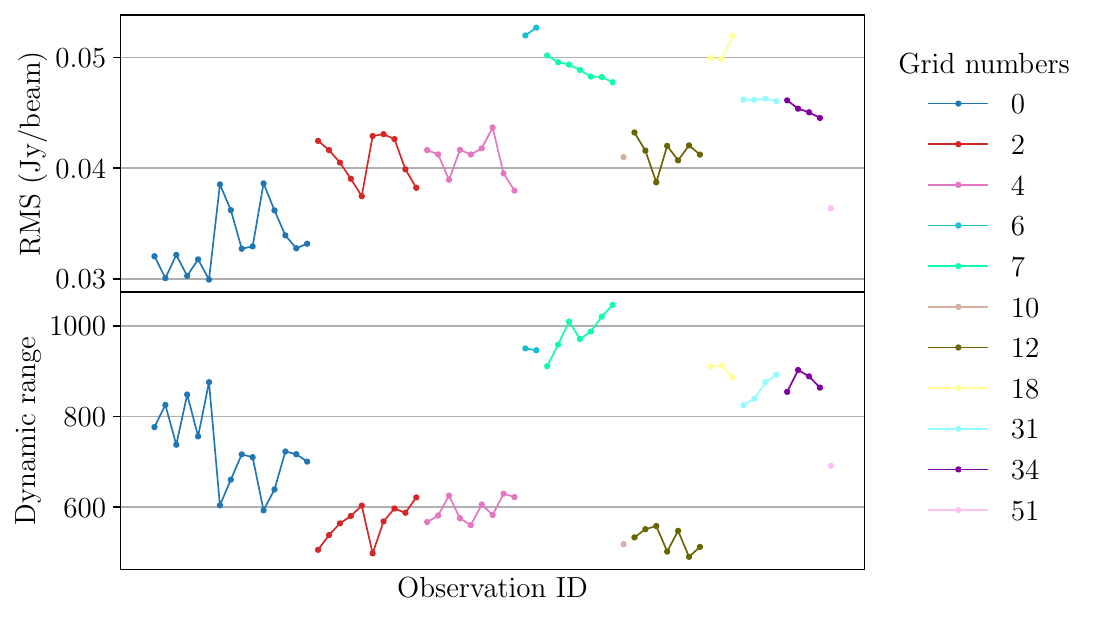}
    \end{center}
    \caption{The RMS of the HERA field at LST 5.2 is shown in the top plot, the dynamic range is shown in the
    bottom plot. Different colours correspond to different grid numbers in the field. Each point
within each pointing corresponds to an observation ID, with observation IDs increasing within a
pointing.}\label{fig:HERA_LST_5.2_image_metrics}
\end{figure}

\begin{figure}[H]
    \begin{center}
        \includegraphics[width=0.95\textwidth]{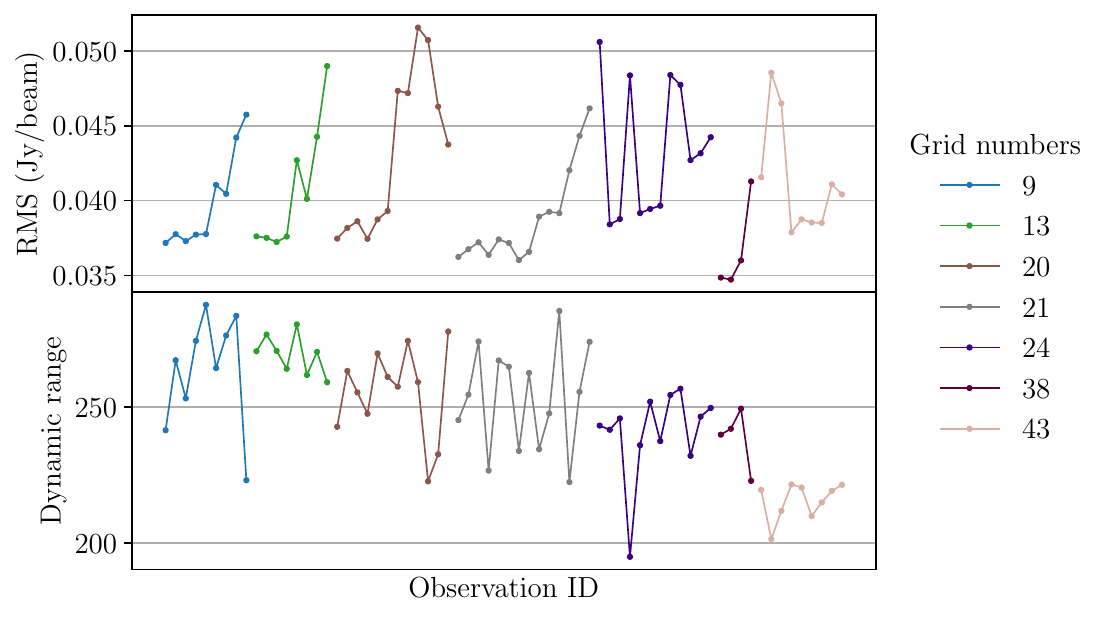}
    \end{center}
    \caption{The RMS of the SKAEOR15 field is shown in the top plot, the dynamic range is shown in the
    bottom plot. Different colours correspond to different grid numbers in the field. Each point
within each pointing corresponds to an observation ID, with observation IDs increasing within a
pointing.}\label{fig:SKAEOR15_image_metrics}
\end{figure}

\raggedbottom{}
\newpage
\section{Amplitude smoothness}
\label{amplitude_smoothness}

This appendix contains figures for the calibration amplitude smoothness metric for the HERA LST 5.2,
SKAEOR5, and SKAEOR15 fields for comparison. Alongside these figures, the HERA LST 2.0 field is
presented to show the clustering behaviour of observations, likely arising from a lack of
observations. This metric indicates smoother amplitude solutions when it is occupies values closer
to 0.

\begin{figure}[H]
    \begin{center}
        \includegraphics[width=0.95\textwidth]{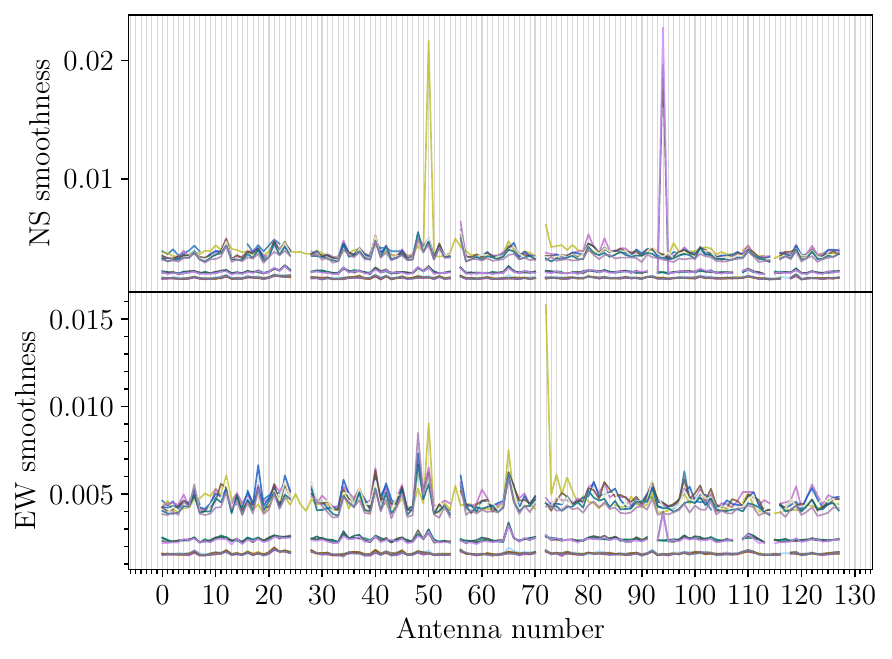}
    \end{center}
    \caption{Smoothness of the NS (top) and EW (bottom) amplitude calibration solutions for
    HERA LST 2.0 for each antenna. Different colours represent a different observation ID. A lower
    value is ideal and indicates smoother calibration amplitude solutions. The clear grouping of
observations likely arise from a lack of data for this field.}\label{fig:HERA_LST_2.0_xx_smoothness}
\end{figure}
\begin{figure}[H]
    \begin{center}
        \includegraphics[width=0.95\textwidth]{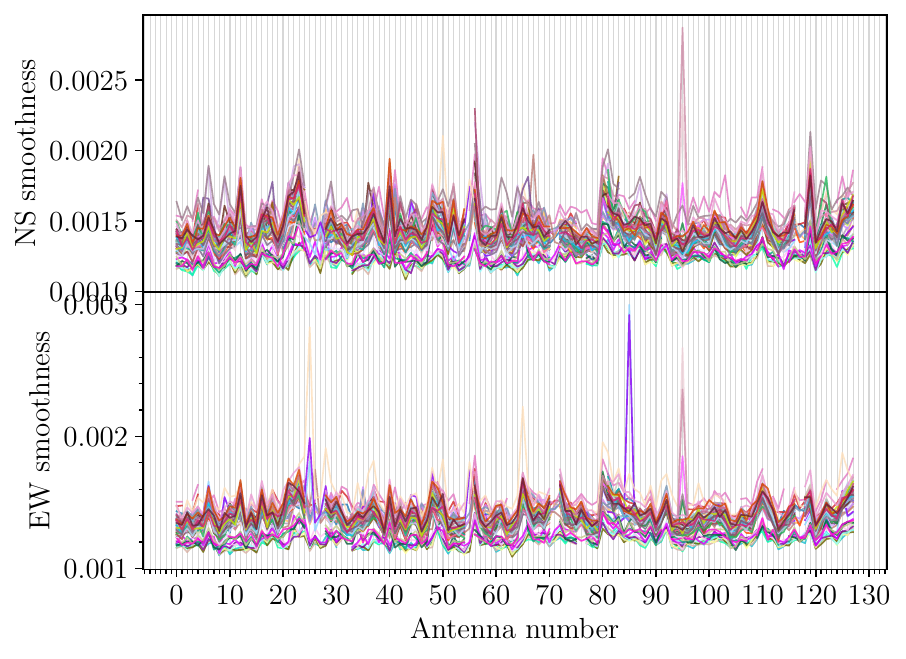}
    \end{center}
    \caption{Smoothness of the NS (top) and EW (bottom) amplitude calibration solutions for
    HERA LST 5.2 for each antenna. Different colours represent a different observation
ID. A lower value is ideal and indicates smoother calibration amplitude
solutions}\label{fig:HERA_LST_5.2_xx_smoothness}
\end{figure}
\begin{figure}[H]
    \begin{center}
        \includegraphics[width=0.95\textwidth]{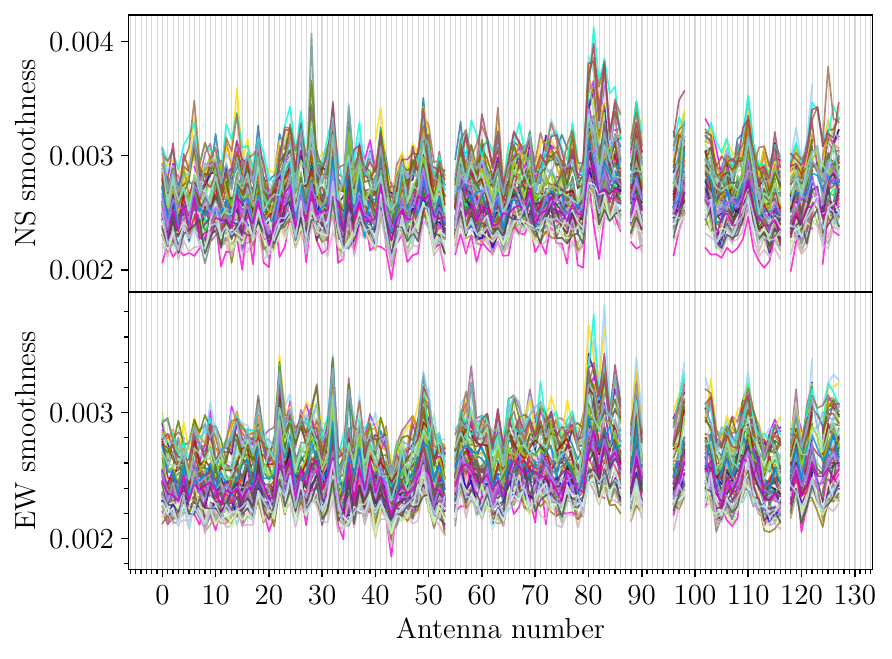}
    \end{center}
    \caption{Smoothness of the NS (top) and EW (bottom) amplitude calibration solutions for
    SKAEOR5 for each antenna. Different colours represent a different observation ID. A lower value
is ideal and indicates smoother calibration amplitude solutions.}\label{fig:SKAEOR5_xx_smoothness}
\end{figure}

\begin{figure}[H]
    \begin{center}
        \includegraphics[width=0.95\textwidth]{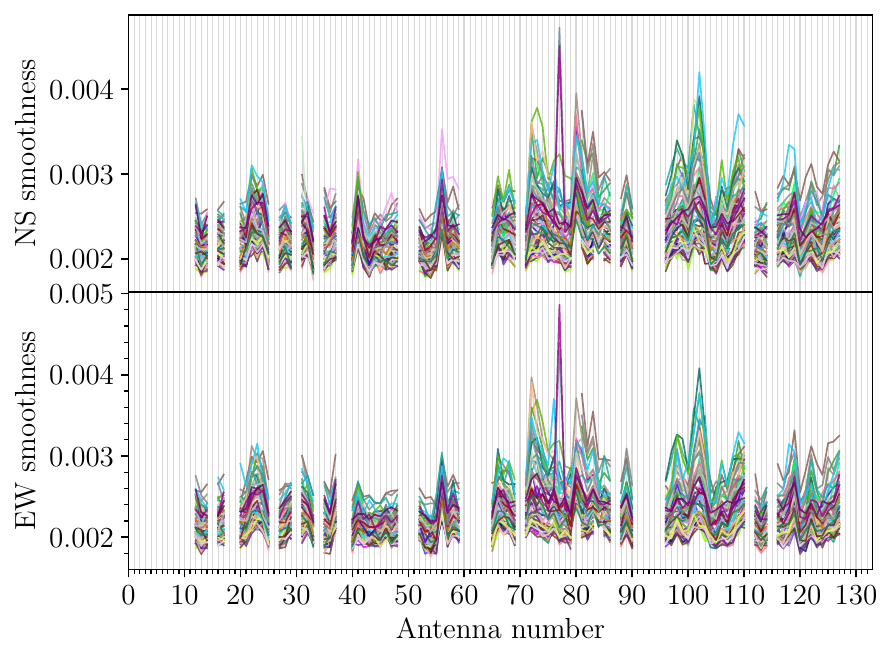}
    \end{center}
    \caption{Smoothness of the NS (top) and EW (bottom) amplitude calibration solutions for
    SKAEOR15 for each antenna. Different colours represent a different observation ID. A lower value
is ideal and indicates smoother calibration amplitude solutions.}\label{fig:SKAEOR15_xx_smoothness}
\end{figure}

\newpage
\section{Smoothness band plots}
\label{smoothness_bands}
This appendix contains figures for the calibration amplitude smoothness metric for the HERA LST 5.2,
SKAEOR5, and SKAEOR15 fields in the form of band plots for comparison. The same data used to
generate the figures in Appendix \ref{amplitude_smoothness} are used here. The plotted red line in
these figures is the median value at each antenna. The shaded region represents the interquartile
range of the data.

\begin{figure}[H]
    \begin{center}
        \includegraphics[width=0.95\textwidth]{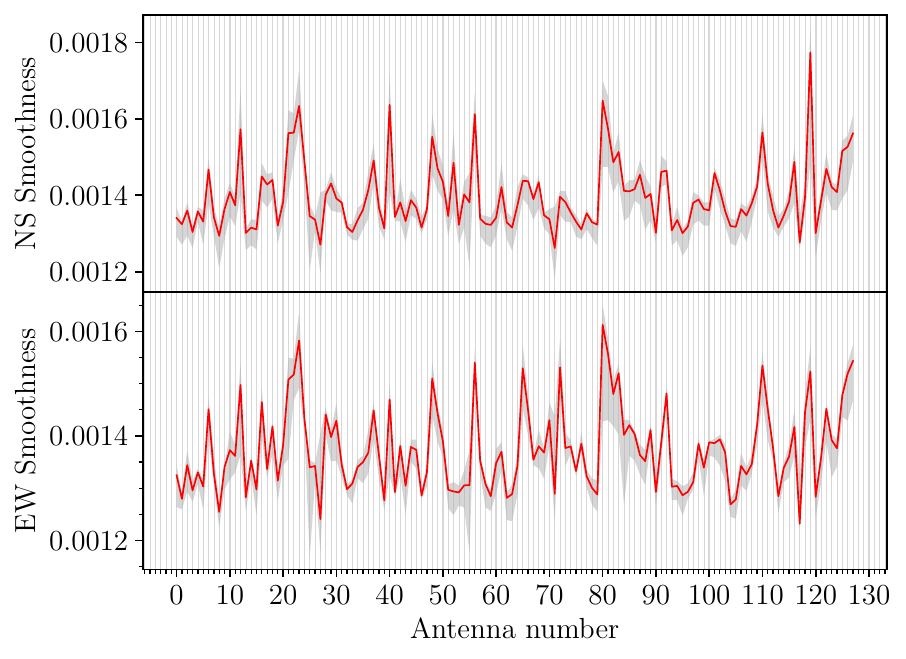}
    \end{center}
    \caption{Smoothness of the NS (top) and EW (bottom) amplitude calibration solutions for
    HERA LST 5.2 for each antenna. The red line represents the median value at each antenna, while
the shaded region represents the interquartile range of the data. A lower value is ideal and indicates
smoother calibration amplitude solutions.}\label{fig:HERA_LST_5.2_xx_smoothness_band}
\end{figure}

\begin{figure}[H]
    \begin{center}
        \includegraphics[width=0.95\textwidth]{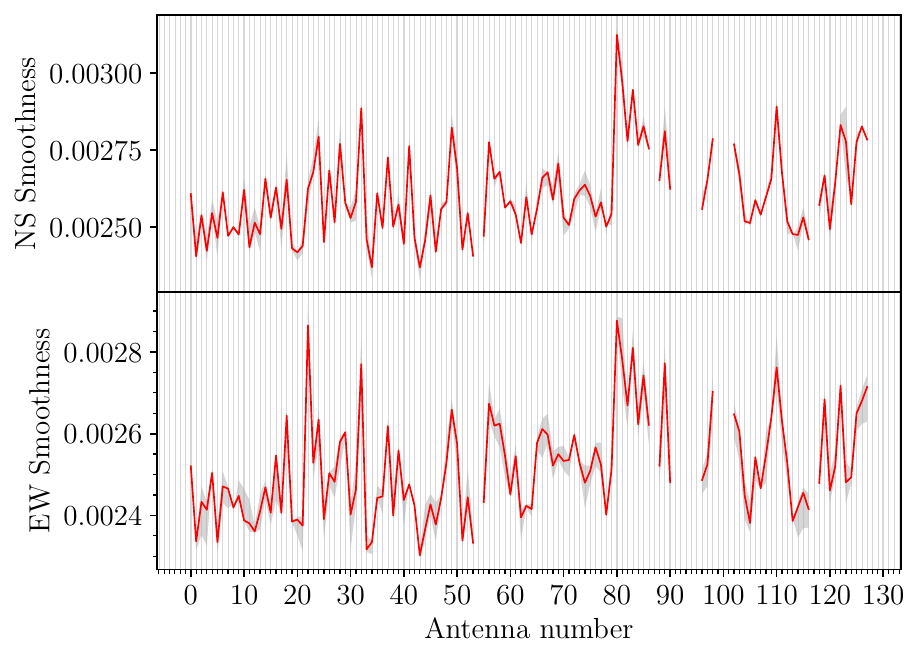}
    \end{center}
    \caption{Smoothness of the NS (top) and EW (bottom) amplitude calibration solutions for
    SKAEOR5 for each antenna. The red line represents the median value at each antenna, while the
shaded region represents the interquartile range of the data. A lower value is ideal and indicates
smoother calibration amplitude solutions.}\label{fig:SKAEOR5_xx_smoothness_band}
\end{figure}

\begin{figure}[H]
    \begin{center}
        \includegraphics[width=0.95\textwidth]{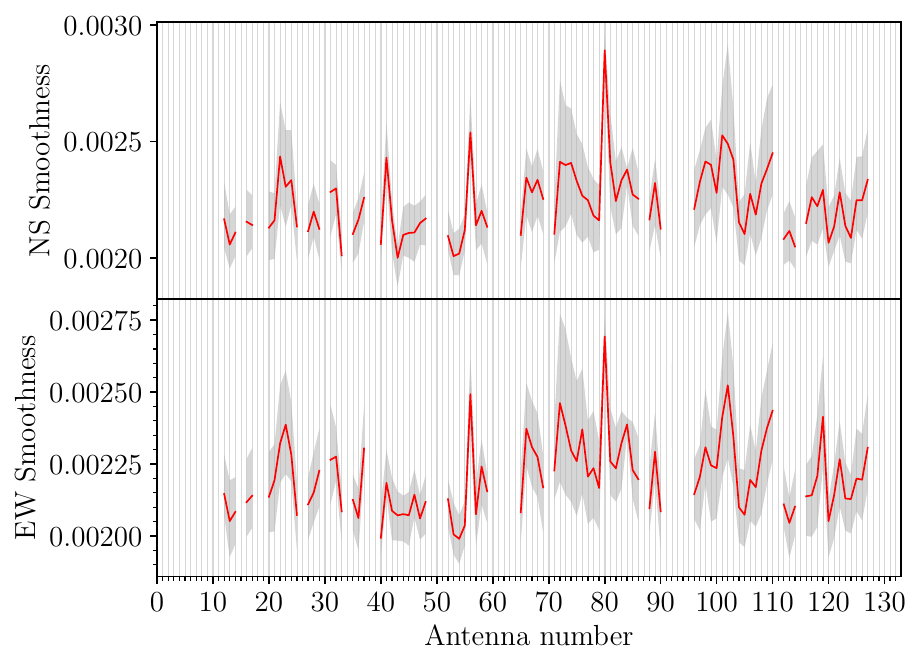}
    \end{center}
    \caption{Smoothness of the NS (top) and EW (bottom) amplitude calibration solutions for
    SKAEOR15 for each antenna. The red line represents the median value at each antenna, while the
shaded region represents the interquartile range of the data. A lower value is ideal and indicates
smoother calibration amplitude solutions.}\label{fig:SKAEOR15_xx_smoothness_band}
\end{figure}

\newpage
\section{RMSE metrics}
\label{rmse_metrics}
This appendix contains figures for the calibration phase RMSE metric for the HERA LST 5.2,
SKAEOR5, and SKAEOR15 fields for comparison. This metric aims to measure the linearity of the phase
solutions, where more linear phase solutions correspond to a lower RMSE metric. 

\begin{figure}
    \begin{center}
        \includegraphics[width=0.95\textwidth]{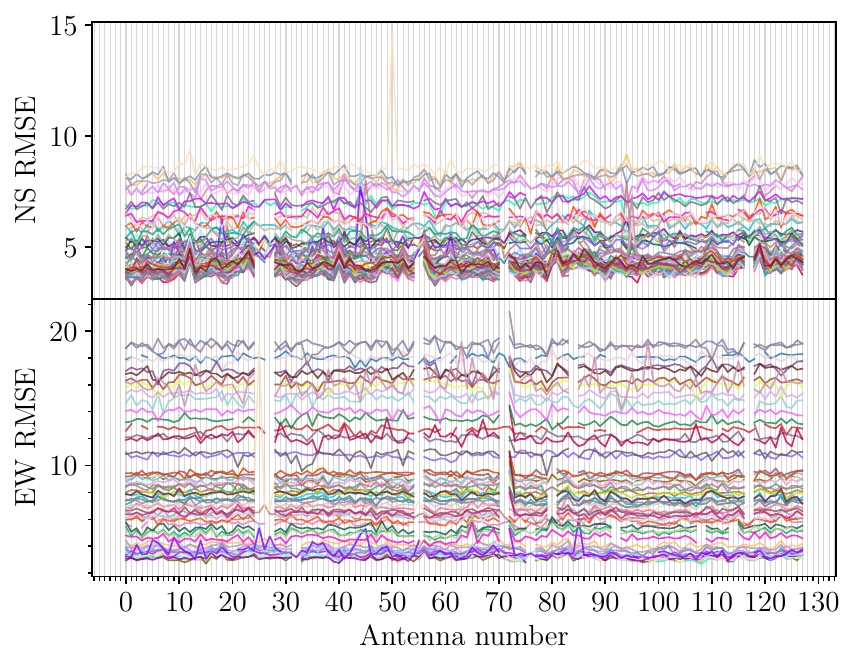} \end{center}
        \caption{RMSE metric for both NS and EW cross polarisations for the HERA LST 5.2 field. Each
        line is a different observation. A value closer to 0 is ideal, and indicates more linear
    phase solutions.}\label{fig:HERA_LST_5.2_rmse}
\end{figure}
\begin{figure}
    \begin{center}
        \includegraphics[width=0.95\textwidth]{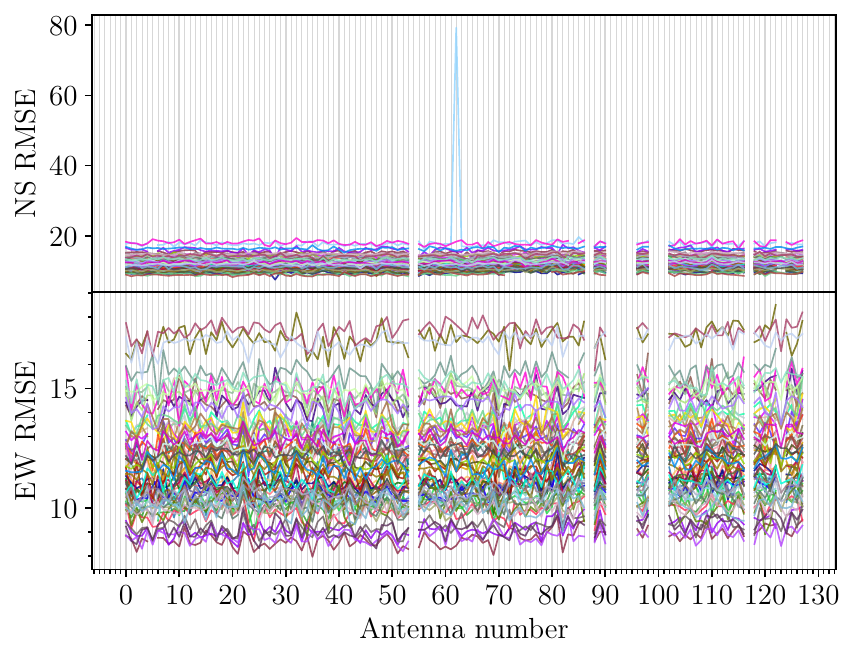}
    \end{center}
    \caption{RMSE metric for both NS and EW cross polarisations for the SKAEOR5 field. Each line is a
    different observation. A value closer to 0 is ideal, and indicates more linear phase
solutions.}\label{fig:SKAEOR5_rmse}
\end{figure}

\begin{figure}
    \begin{center}
        \includegraphics[width=0.95\textwidth]{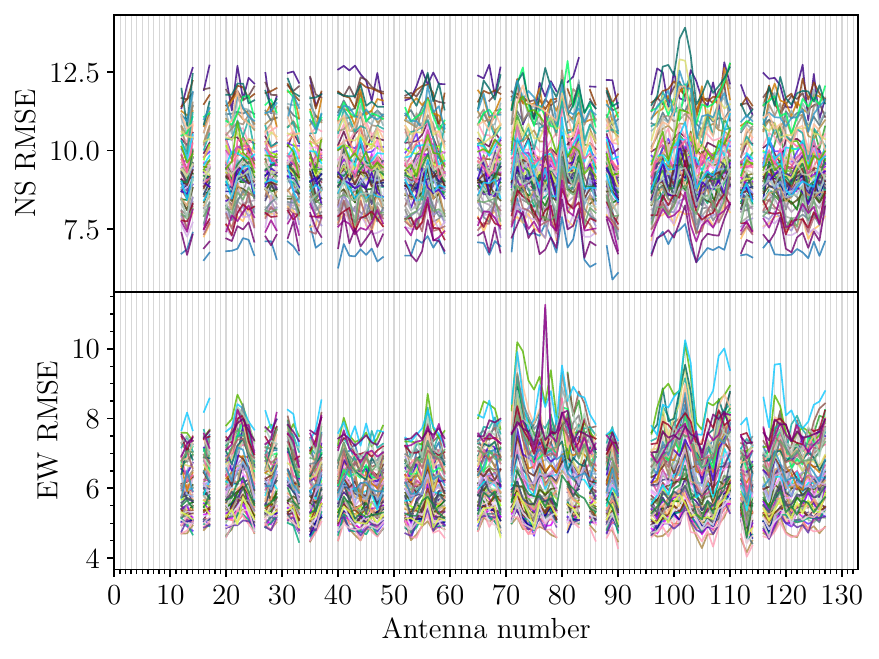}
    \end{center}
    \caption{RMSE metric for both NS and EW cross polarisations for the SKAEOR15 field. Each line
        represents a different observation. A value closer to 0 is ideal, and indicates more linear
    phase solutions.}\label{fig:SKAEOR15_rmse}
\end{figure}

\raggedbottom{}
\newpage
\section{Average Euclidean distance metric}
\label{euclid}
This appendix contains figures for the calibration phase average Euclidean distance metric for the
HERA LST 5.2, SKAEOR5, and SKAEOR15 fields.

\begin{figure}[H]
    \begin{center}
        \includegraphics[width=0.95\textwidth]{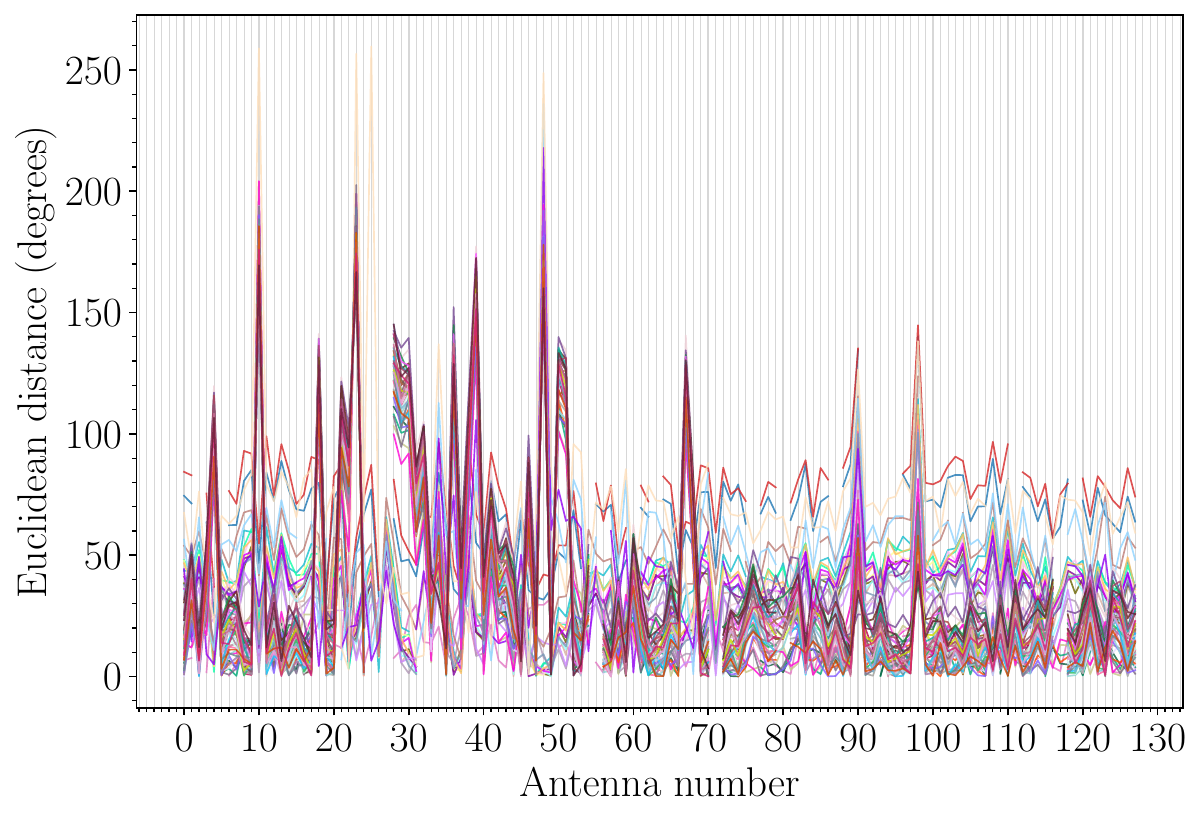}
    \end{center}
    \caption{The average Euclidean distance between the NS and EW cross polarisations for each
    antenna in each observation for the HERA LST 5.2 field. A value closer to 0 is ideal and signals
that the solutions are more similar.}\label{fig:HERA_LST_5.2_euclid}
\end{figure}
\begin{figure}[H]
    \begin{center}
        \includegraphics[width=0.95\textwidth]{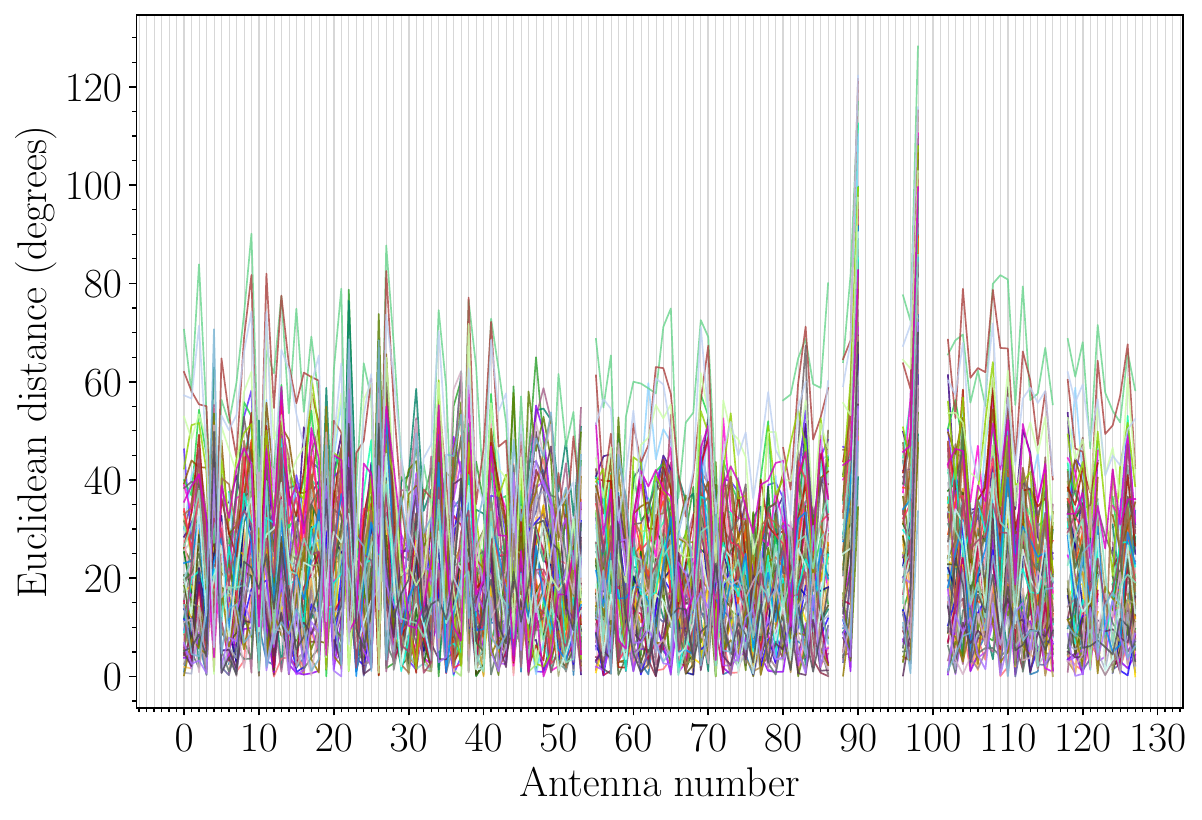}
    \end{center}
    \caption{The Euclidean distance between the NS and EW cross polarisations for each antenna in
    each observation for the SKAEOR5 field. A value closer to 0 is ideal and signals that the
solutions are more similar.}\label{fig:SKAEOR5_euclid}
\end{figure}

\begin{figure}[H]
    \begin{center}
        \includegraphics[width=0.95\textwidth]{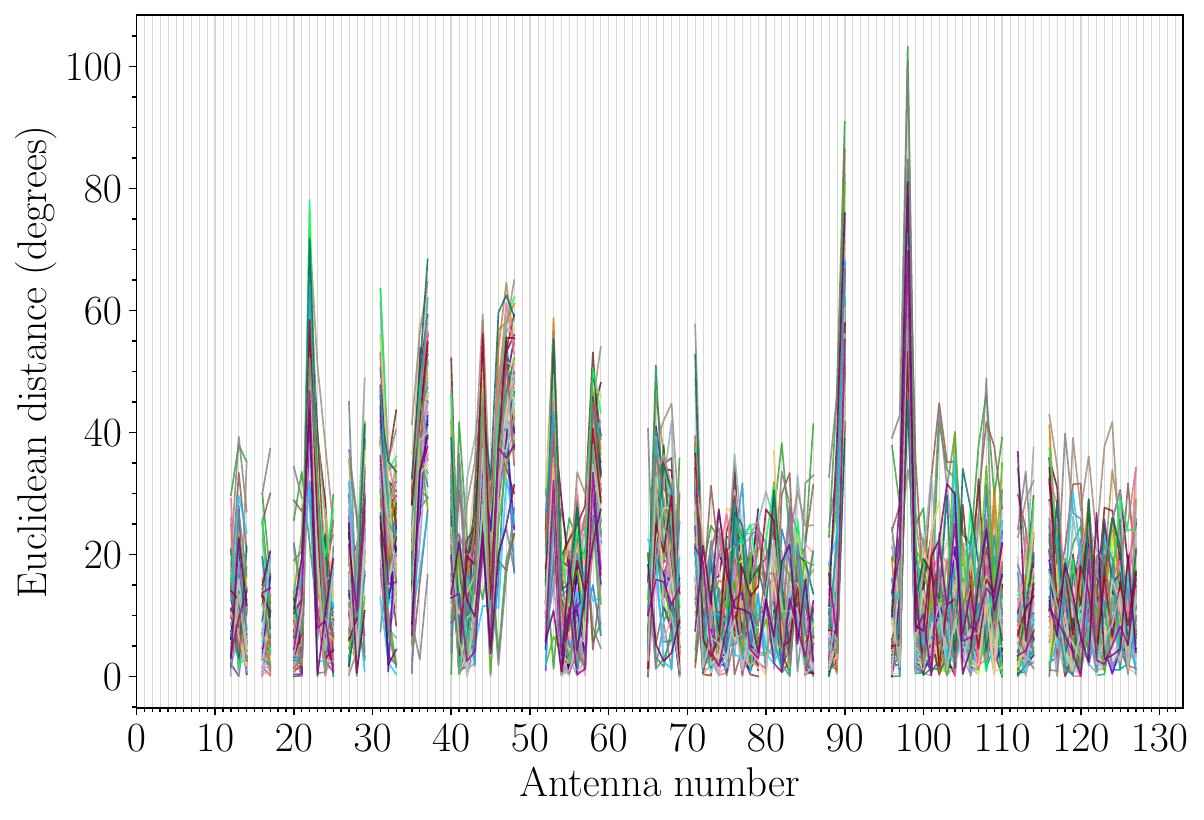}
    \end{center}
    \caption{The Euclidean distance between the NS and EW cross polarisations for each antenna in
    each observation for the SKAEOR15 field. A value closer to 0 is ideal and signals that the
solutions are more similar.}\label{fig:SKAEOR15_euclid}
\end{figure}

\end{document}